\documentclass[aps,prd,preprint,groupedaddress]{revtex4}
\begin{document}

%\preprint{}

\title{Light-flavor sea-quark distributions in the nucleon\\
in the SU(3) chiral quark soliton model (II) \\
--- theoretical formalism ---}


\author{M.~Wakamatsu}
\email[]{wakamatu@miho.rcnp.osaka-u.ac.jp}
%\thanks{}
\affiliation{Department of Physics, Faculty of Science, \\
Osaka University, \\
Toyonaka, Osaka 560, JAPAN}


%\date{\today}

\begin{abstract}
The path integral formulation is given to obtain quark and antiquark
distribution functions in the nucleon within the flavor SU(3) version of 
the chiral quark soliton model. The basic model action is a straightforward 
generalization of the corresponding SU(2) one, except for one distinguishable
feature, i.e. the presence of the SU(3) symmetry breaking term arising from 
the sizably large mass difference $\Delta m_s$ between the strange and 
nonstrange quarks. We treat this SU(3) symmetry breaking effect by relying 
upon the first order perturbation theory in the mass parameter $\Delta m_s$.
We also address to the problem of ordering ambiguity of the relevant 
collective space operators, which arises in the evaluation of the parton 
distribution functions at the subleading order of $1/N_c$ expansion.
\end{abstract}

%\keywords{}

\maketitle

%\section{}
%\subsection{}
%\subsubsection{}

\section{Introduction}

In the preceding paper \cite{W03}, which is referred to as I, we have shown
that the flavor SU(3) version of the chiral quark soliton model (CQSM)
can give reasonable predictions for the hidden strange quark distributions
in the nucleon, while preserving the success of the SU(2) CQSM.
The detailed theoretical formulation of the model was left out, however,
in consideration of its quite elaborate nature.
The purpose of the present paper is to make up for this point.

The generalization of the CQSM to the case of flavor SU(3) was already
done many years ago independently by two groups \cite{WAR92},\cite{BDGPPP93}.
The basic dynamical assumption of the SU(3) CQSM
is very similar to that of the SU(3) Skyrme model \cite{G84},\cite{MNP84}.
It is the embedding of the SU(3) hedgehog mean-field into
the SU(3) matrix followed by the quantization of the collective
rotational motion in the full SU(3) collective coordinate space.
The physical octet and decuplet baryons including the nucleon with
good spin and flavor quantum numbers are obtained through this
quantization process.
For the usual low energy observables of baryons like the magnetic
moments or the axial-vector couplings, the theory can be formulated
by using the standard cranking procedure which is familiar in the
nuclear theory of collective rotation. However, what we want to
investigate here is not the usual low energy observables of baryons
but the quark and antiquark distributions in the nucleon, which are
fully relativistic objects. For obtaining these quantities, we must
evaluate nucleon matrix elements of quark bilinear operators containing
two space-time coordinates with light-cone separation.
The most convenient method for investigating such quantities is the
path integral formalism, which was already used in the formulation
of the similar observables in the SU(2) version of the
CQSM \cite{DPPPW96}\nocite{DPPPW97}\nocite{WK98}\nocite{WK99}\nocite{DPP88}\nocite{WY91}--\cite{W92}.

The standard mean-field approximation in the nuclear theory corresponds
to the stationary-phase approximation in the path integral
formalism \cite{DPP88}. The rotational motion of the symmetry
breaking mean-field configuration, which appears as a zero-energy mode,
is treated by using the first order
perturbation theory in the collective rotational velocity $\Omega$ of
the soliton. This is justified since the velocity of this collective
rotational motion is expected to be much slower than the velocity of
intrinsic quark motion in the hedgehog mean field.
According to this theoretical structure of the model, any baryon
observables including parton distribution functions (PDF) are given
as a sum of the $O(\Omega^0)$ contributions and the $O(\Omega^1)$
one \cite{DPP88},\cite{WY91}.

A completely new feature of the SU(3) CQSM, that is not shared by the
SU(2) model, is the existence of SU(3) symmetry breaking term due to
the appreciable mass difference between the strange and nonstrange
quarks. We believe that this mass difference (or the mass of the strange
quark itself) of the order 100 MeV is still much smaller than the
typical energy  scale of hadron physics of the order 1 GeV. and it
can be treated by relying upon the perturbation theory.

Now, in the next section, we start to explain the detailed path
integral formulation of the SU(3) CQSM for evaluating PDF.
After explaining the general theoretical structure of the model,
we shall discuss the $O(\Omega^0)$ contributions to the PDF, the
$O(\Omega^1)$ contributions and the first order corrections
in $\Delta m_s$ in three separate subsections. Finally, in sect.4,
we briefly summarize our achievement as well as what still remain to
be clarified in future studies.

%\vspace{4mm}
\section{Formulation of the model}

We start with the familiar definition of the quark distribution 
function given as \cite{CS82}
\begin{equation}
   q (x) \ = \ \frac{1}{4 \pi} \,\,\int_{-\infty}^\infty
   \,\,d z_0 \,\,e^{\,i \,x \,M_N \,z_0} \,\,
   {\langle N (\mbox{\boldmath $P$} = 0) \,| \, \psi^\dagger (0) \,
   O_a \,\psi(z) \,| \,
   N (\mbox{\boldmath $P$} = 0) \rangle} \,\, 
   {|}_{z_3 = - z_0, \,z_\perp = 0} \,\, .
\end{equation}
Here $O_a$ is to be taken as 
\begin{equation}
 O_a = \ \lambda_a \,(1 + \gamma^0 \gamma^3) , \label{qextend}
\end{equation}
with $a = 0, 3$, and $8$ for unpolarized
distribution functions (note here we take that $\lambda_0 = 1$), while
\begin{equation}
 O_a = \ \lambda_a \,(1 + \gamma^0 \gamma^3) \,\gamma_5 ,
 \label{dqextend}
\end{equation}
for longitudinally polarized ones. We recall that the above 
definition of the quark distribution function can formally be 
extended to the negative $x$ region. The function $q(x)$ with 
negative argument should actually be interpreted as giving an 
antiquark distribution with physical value of $x \,( > 0)$
according to the rule :
\begin{equation}
 q(-x) = - \,\bar{q} (x) \hspace{24mm} (0 < x < 1) ,
\end{equation}
for the unpolarized distributions, and
\begin{equation}
 \Delta q(-x) = + \,\Delta \bar{q} (x)\hspace{20mm} (0 < x < 1) ,
\end{equation}
for the longitudinally polarized distributions. Here, the sign
difference between the two types of distributions arises from the 
different ways of their transformations under charge 
conjugation.

As was explained in the previous paper, the startingpoint 
of our theoretical analysis is the following path integral 
representation of a matrix element of a bilocal and bilinear quark 
operator between the nucleon state with definite momentum : 
\begin{eqnarray}
   &\,& \langle N (\mbox{\boldmath $P$}) \,| \,\psi^\dagger (0) \,O_a
   \,\psi(z) \,| \,N (\mbox{\boldmath $P$})\rangle
   \ \ = \ \ \frac{1}{Z} \,\,\int \,\,d^3 x  \,\,d^3 y \,\,
   e^{\,- \,i \mbox{\boldmath $P$} \cdot \mbox{\boldmath $x$}} \,\,
   e^{\,i \,\mbox{\boldmath $P$} \cdot \mbox{\boldmath $y$}} \,\,
   \int {\cal D} U \nonumber \\
   &\times& \! \int {\cal D}
   \psi \,\,{\cal D} \psi^\dagger \,\,
   J_N (\frac{T}{2}, \mbox{\boldmath $x$}) \,\,\psi^\dagger (0) \,
   O_a \,\psi(z) \,\,J_N^\dagger (-\frac{T}{2},
   \mbox{\boldmath $y$}) \,\,
   \exp \,[\,\,i \int \,d^4 x \,\, {\cal L} (x) 
   \, ] \, , \ \ \ \ \ 
\end{eqnarray}
where
\begin{eqnarray}
   {\cal L} \ \ = \ \ \bar{\psi} \,(\,i \not\!\partial \ - \ 
   M U^{\gamma_5} (x) - \Delta m_s P_s \,) \,\psi \,\, ,
\end{eqnarray}
with $U^{\gamma_5} (x) = \mbox{exp} [ i \gamma_5 \lambda_a \pi_a (x) 
/ f_{\pi} ]$ being the basic lagrangian of the CQSM with three
flavors \cite{WAR92},\cite{BDGPPP93}.
Here, the mass difference $\Delta m_s$ between the
strange quark and nonstrange quarks is introduced with use of the 
projection operator
\begin{equation}
 P_s \ = \ \frac{1}{3} - \frac{1}{\sqrt{3}} \lambda_8 \ = \ 
 \left( \begin{array}{ccc}
 \ 0 \ & \ 0 \ & \ 0 \ \\
 \ 0 \ & \ 0 \ & \ 0 \ \\
 \ 0 \ & \ 0 \ & \ 1 \ 
 \end{array}
 \right)
\end{equation}
for the $s$-quark component. The quantity
\begin{equation}
   J_N (x) \ \ = \ \ \frac{1}{N_c !} \,\, 
   \epsilon^{\alpha_1 \cdots \alpha_{N_c}} \,\,
   \Gamma_{Y T T_3 ; J J_3}^{\{f_1 \cdots f_{N_c}\}} \,\,
   \psi_{\alpha_1 f_1} (x) \cdots 
   \psi_{\alpha_{N_c} f_{N_c}} (x) \,\, ,
\end{equation}
is a composite operator carrying the quantum numbers 
$Y T T_3 , J J_3$ (hypercharge, isospin and spin) of the baryon, where
$\alpha_i$ the color index, while 
$\Gamma_{Y T T_3 ; J J_3}^{\{f_1 \cdots f_{N_c}\}}$ is a symmetric 
matrix in spin flavor indices $f_i$. A basic dynamical assumption 
of the SU(3) CQSM (which one may notice is similar to that of
the SU(3) Skyrme model \cite{G84}) is the embedding of the
SU(2) self-consistent 
mean-field solution of hedgehog shape into the SU(3) matrix as
\begin{equation}
 U_0^{\gamma_5} (\mbox{\boldmath $x$}) = 
 \left( \begin{array}{cc}
 e^{i \gamma_5 \mbox{\boldmath $\tau$}
 \cdot \hat{\mbox{\boldmath $r$}} F(r)} & 0 \\
 0 & 1
 \end{array}
 \right) . \label{hedgehog}
\end{equation}
That this would give the lowest energy classical configuration can 
be deduced from a simple variational argument \cite{W02P}.
In fact, an arbitrary small variation of the $(3,3)$ component of
$U^{\gamma_5}_0 (\mbox{\boldmath $x$})$ would induce a change of the
strange-quark single-particle spectra in such a way that weak bound
states appear from the positive energy Dirac continuum as well as from
the negative energy 
one in a charge-conjugation symmetric way. Since only the negative
energy continuum is originally occupied, this necessarily increases
the total energy of the baryon-number-one system.
Because of energy-degeneracy of all the configurations attainable from
the above configuration under the spatial rotation or the rotation 
in the flavor SU(3) internal space, a spontaneous zero-energy 
rotational mode necessarily arises. We also notice the existence 
of another important zero mode corresponding to the translational 
motion of the soliton center.
As in the previous paper \cite{DPPPW96}--\cite{WK99}, the 
translational zero-mode is treated by using an approximate momentum 
projection procedure (of the nucleon state), which amounts to 
integrating over all the shift $\mbox{\boldmath $R$}$ of the soliton 
center-of-mass coordinates :
\begin{equation}
   \langle N (\mbox{\boldmath $P$}) \,| \,
   \psi^\dagger (0) \,O_a \,\psi (z) \,| \,N
   (\mbox{\boldmath $P$}) \rangle
   \ \longrightarrow \ \int \,d^3 R \,\,
   \langle N (\mbox{\boldmath $P$}) \,| \,
   \psi^\dagger (0,- \mbox{\boldmath $R$}) \,
   O_a \,\psi (z_0,\mbox{\boldmath $z$} - 
   \mbox{\boldmath $R$}) \,| \,N
   (\mbox{\boldmath $P$}) \rangle \, .
\end{equation}
On the other hand, the rotational zero modes can be treated by 
introducing a rotating meson field of the form
\begin{eqnarray}
   U^{\gamma_5} ( \mbox{\boldmath $x$}, t) \ \ = \ \ A(t) \,\,
   U_0^{\gamma_5} (\mbox{\boldmath $x$}) \,\,A^\dagger (t) \,\, ,
\end{eqnarray}
where $A(t)$ is a time-dependent SU(3) matrix in flavor space.
A key identity in the following manipulation is as follows,
\begin{equation}
 \bar{\psi} \,(\,i \! \not\!\partial \ - \ 
 M U^{\gamma_5} (x) - \Delta m_s P_s \,) \,\psi \ = \ 
 \psi_A^\dagger ( i \partial_t - H - \Delta H - \Omega) \psi_A
\end{equation}
where
\begin{eqnarray}
   \psi_A &=& A^\dagger (t) \,\psi \,\, , \hspace{5mm} \\
   H &=& \frac{\mbox{\boldmath $\alpha$} \cdot \nabla}{i} 
   \ + \ M \,\beta \,U^{\gamma_5}_0 (\mbox{\boldmath $x$})
   \,\, ,\\
   \Delta H &=& \Delta m_s \cdot \gamma^0 A^\dagger (t) 
   \left( \frac{1}{3} - \frac{1}{\sqrt{3}} \lambda_8 \right)
   A(t) , \\
   \Omega &=& - \,i \,A^\dagger (t) \,\dot{A} (t) .
\end{eqnarray}
Here $H$ is a static Dirac Hamiltonian with the background pion 
field $U^{\gamma_5}_0 (\mbox{\boldmath $x$})$,
playing the role of mean-field potential 
for quarks, whereas $\Delta H$ is the SU(3) symmetry breaking
correction to $H$. The quantity $\Omega$ is 
the SU(3)-valued angular velocity matrix later to be quantized in 
an appropriate way. At this stage, it is convenient to introduce 
a change of quark field variable $\psi \rightarrow \psi_A$, which 
amounts to getting on a body-fixed rotating frame of a soliton.
Denoting $\psi_A$ anew $\psi$ for notational simplicity, the 
nucleon matrix element (8) can then be written as
\begin{eqnarray}
   &\,& \langle N (\mbox{\boldmath $P$}) \,| \,\psi^\dagger (0) \,O_a
   \,\psi(z) \, | \,N (\mbox{\boldmath $P$}) \rangle \nonumber \\
   &=& \frac{1}{Z} \,\,\Gamma^{\{ f \}} \,\,{\Gamma^{\{ g \}}}^* \,\,
   \int \,\,d^3 x \,\,d^3 y \,\,
   e^{-i \mbox{\boldmath $P$} \cdot \mbox{\boldmath $x$}} \,\,
   e^{i \mbox{\boldmath $P$} \cdot \mbox{\boldmath $y$}} \,\,
   \int \,\,d^3 R \nonumber \\
   &\times& \int \,\,{\cal D} A \,\,{\cal D} \psi \,\,
   {\cal D} \psi^\dagger
   \,\,\exp \,[ \,\,i \,\int \,d^4 x \,\,\psi^\dagger 
   (\,i \partial_t - H - \Delta H - \Omega) \,\psi \,] \,\,
   \prod^{N_c}_{i = 1} \,\,\,
   [ \,A ( \frac{T}{2}) \,\,\psi_{f_i}
   ( \frac{T}{2}, \mbox{\boldmath $x$} ) \,] \nonumber \\
   &\times& \psi^\dagger (0, - \mbox{\boldmath $R$})
   \,\,A^\dagger (0) \,O_a \,A(z_0) \,\,
   \psi(z_0,\mbox{\boldmath $z$} - \mbox{\boldmath $R$}) \,\,
   \prod^{N_c}_{j = 1} \,\,\, [ \,\psi_{g_j}^\dagger
   (- \frac{T}{2}, \mbox{\boldmath $y$}) \,\,
   A^\dagger (-\frac{T}{2})] \,\, . \label{base1}
\end{eqnarray}
Performing the path integral over the quark fields, 
we obtain
\begin{eqnarray}
   &\,& \langle N (\mbox{\boldmath $P$}) \,
   | \,\psi^\dagger (0) \,O_a \,\psi (z) \,
   | \,N (\mbox{\boldmath $P$}) \rangle  \nonumber \\
   &=& \frac{1}{Z} \,\,\tilde{\Gamma}^{\{ f \}} \,\,
   {\tilde{\Gamma}}^{{\{ g \}}^\dagger} \,\,N_c \,\,
   \int \,\,d^3 x \,\,d^3 y \,\,e^{\,-i \,\mbox{\boldmath $P$}
   \cdot \mbox{\boldmath $x$}}
   \,\,e^{\,i \mbox{\boldmath $P$} \cdot \mbox{\boldmath $y$}} \,\,
   \int d^3 R \,\,\int {\cal D} A \nonumber \\
   &\times& \!\! \Bigl\{ \, 
   {}_{f_1} \langle\frac{T}{2}, \mbox{\boldmath $x$}
   \,| \,\frac{i}
   {\,i \partial_t - H - \Delta H - \Omega} \,| \,0,- 
   \mbox{\boldmath $R$}
   \rangle_{\gamma} \,\,
   {( A^\dagger (0) O_a A(z_0) )}_{\gamma \delta} \nonumber \\
   &\,& \hspace{50mm} \times \,
   {}_\delta \langle z_0, \mbox{\boldmath $z$} -
   \mbox{\boldmath $R$} \,| \,\frac{i}
   {\,i \partial_t - H - \Delta H - \Omega} \,
   | - \frac{T}{2}, \mbox{\boldmath $y$} \rangle_{g_1} \nonumber \\
   &-& \!
   \mbox{Tr} \,{\bigl( \,\langle z_0, \mbox{\boldmath $z$} -
   \mbox{\boldmath $R$} \,| \,
   \frac{i}{i \partial_t - H - \Delta H - \Omega} \,| \,
   0, - \mbox{\boldmath $R$} \rangle
   \,\,A^\dagger (0) O_a A (z_0) \,\bigr)} \nonumber \\
   &\,& \hspace{55mm} \times \,
   {}_{f_1} \langle \frac{T}{2},\mbox{\boldmath $x$} \,| \,
   \frac{i}{\,i \partial_t - H - \Delta H - \Omega} \,| - \frac{T}{2},
   \mbox{\boldmath $y$}\rangle_{g_1} \,\, \Bigr\} \nonumber \\
   &\times& \prod^{N_c}_{j = 2} \,\,\,[ \,{}_{f_j}
   \langle \frac{T}{2}, \mbox{\boldmath $x$}
   \,| \,\frac{i}{\,i \partial_t - H - \Delta H - \Omega} \,
   | -\frac{T}{2}, \mbox{\boldmath $y$}\rangle_{g_j} \,]
   \nonumber \\
   &\,& \hspace{55mm} \times \,
   \exp \,[\,N_c \,\mbox{Sp} \log \,
   (\,i \partial_t - H - \Delta H \Omega) \, ] \,\, , \ \ \ \ \ 
   \label{base2}
\end{eqnarray}
with $\tilde{\Gamma}^{ \{ f \}} 
= \Gamma^{ \{f \}} [ A (T / 2) ]^{N_c}$ etc. Here $\mbox{Tr}$ is to
be taken over spin-flavor indices.
Now the strategy of the following manipulation is in order.
As in all the previous works, we assume that the collective 
rotational velocity of the soliton is much slower than the velocity 
of internal quark motion, which provides us with a theoretical
support to a perturbative treatment in $\Omega$.
Since $\Omega$ is known to be an $O(1 / N_c)$ quantity, 
this perturbative expansion in $\Omega$ can also be taken as a 
$1 / N_c$ expansion. We shall retain terms up to the first order 
in $\Omega$. We also use the perturbative expansion in 
$\Delta m_s$, which is believed to be a small parameter as compared
with the typical energy scale of low energy
QCD $( \sim 1 \mbox{GeV})$.

Applying this expansion to (\ref{base2}), we obtain
\begin{eqnarray}
 \langle N (\mbox{\boldmath $P$}) \vert \psi^\dagger (0) O_a
 \psi(z) \vert N(\mbox{\boldmath $P$}) \rangle &=& 
 {\langle N (\mbox{\boldmath $P$}) \vert \psi^\dagger (0) O_a
 \psi(z) \vert N(\mbox{\boldmath $P$}) \rangle}^{\Omega^0} 
 \nonumber \\
 &+& {\langle N (\mbox{\boldmath $P$}) \vert \psi^\dagger (0) O_a
 \psi(z) \vert N(\mbox{\boldmath $P$}) \rangle}^{\Omega^1}
 \nonumber \\
 &+& {\langle N (\mbox{\boldmath $P$}) \vert \psi^\dagger (0) O_a
 \psi(z) \vert N(\mbox{\boldmath $P$}) \rangle}^{\Delta m_s} 
 \ + \ \cdots \ \ .
\end{eqnarray}
To be more explicit, they are given by
\begin{eqnarray}
   &\,& \!\!\!\!\! {\langle N (\mbox{\boldmath $P$}) 
   \,| \,\psi^\dagger (0) \,
   O_a \,\psi (z) \,| \,
   N (\mbox{\boldmath $P$}) \rangle}^{\Omega^0}  \nonumber \\
   &=& \frac{1}{Z} \,\,{\,\tilde{\Gamma}}^{\{ f \}} \,\,
   {\tilde{\Gamma}}^{{\{ g \}}^\dagger} \,\,N_c \,\, 
   \int \,\,d^3 x \,\,d^3 y  \,\,e^{\,-i \mbox{\boldmath $P$} 
   \cdot \mbox{\boldmath $x$}} 
   \,\,e^{\,i \mbox{\boldmath $P$} \cdot \mbox{\boldmath $y$}}
   \,\,\int \,\,d^3 R \,\,\int \,\,{\cal D} A \,\,
   {(\tilde{O}_a)}_{\gamma \delta} \nonumber \\
   &\times& \! \Bigl[ \,{}_{f_1}
   \langle\frac{T}{2}, \mbox{\boldmath $x$} \,| \, 
   \frac{i}{i \partial_t - H} \,| \,
   0, -\mbox{\boldmath $R$} \rangle_{\gamma} \cdot
   {}_ {\delta} \langle z_0, 
   \mbox{\boldmath $z$} - \mbox{\boldmath $R$}
   \,| \,\frac{i}{i \partial_t - H} \,|\, - 
   \frac{T}{2}, \mbox{\boldmath $y$} \rangle {}_{g_1} \nonumber \\
   &-& \! {}_{\delta} \langle z_0, 
   \mbox{\boldmath $z$} - \mbox{\boldmath $R$} \,| \,
   \frac{i}{i \partial_t - H} \,| \,
   0, - \mbox{\boldmath $R$} \rangle_{\gamma}
   \cdot {}_{f_1} \langle \frac{T}{2}, \mbox{\boldmath $x$} \,| \,
   \frac{i}{i \partial_t - H} \,| \,- \frac{T}{2}, 
   \mbox{\boldmath $y$} \rangle {}_{g_1} \,\Bigr]  \nonumber \\
   &\times& \! \prod^{N_c}_{j = 2} \,\,\,\bigl[\, {}_{f_j}
   \langle \frac{T}{2}, \mbox{\boldmath $x$}
   \,| \,\frac{i}{i \partial_t - H} \,| \,
   - \frac{T}{2}, \mbox{\boldmath $y$} \rangle_{g_j} \,\bigr] 
   \,\cdot \,
   \exp \,[\,N_c \,\mbox{Sp} \log (\,i \partial_t - H)
   \,+ \,i \,\frac{I}{2} 
   \,\int \Omega_a^2 \,d t \,\,]  \, ,
\end{eqnarray}
\begin{eqnarray}
   &\,& \langle N (\mbox{\boldmath $P$}) \,| \,\psi^\dagger (0) \,
   O_a \,\psi (z) \,| \,
   N (\mbox{\boldmath $P$}) \rangle^{\Omega^1}  \nonumber \\
   &=& \frac{1}{Z} \,\,{\,\tilde{\Gamma}}^{\{ f \}} \,\,
   {\tilde{\Gamma}}^{{\{ g \}}^\dagger} \,\,N_c \,\, 
   \int \,\,d^3 x \,\,d^3 y  \,\,e^{\,-i \mbox{\boldmath $P$} 
   \cdot \mbox{\boldmath $x$}} 
   \,\,e^{\,i \mbox{\boldmath $P$} \cdot \mbox{\boldmath $y$}}
   \,\,\int \,\,d^3 R \,\,\int \,\,{\cal D} {\cal A} \nonumber \\
   &\times& \! \Biggl\{ \,
   \int d^3 z^{\prime} \,\,d z_0^\prime \,\,\,\,
   i \,\Omega_{\alpha \beta} (z_0^\prime) \,\,
   {(A^\dagger (0) O_a A(z_0))}_{\gamma \delta} \nonumber \\
   &\times& \!\!
   \Bigl[ \,{}_{f_1}
   \langle\frac{T}{2}, \mbox{\boldmath $x$} \,| \, 
   \frac{i}{i \partial_t - H} \,| \,
   z_0^\prime, \mbox{\boldmath $z$}^\prime \rangle_{\alpha} \cdot
   {}_ {\beta} \langle z_0^\prime, \mbox{\boldmath $z$}^\prime 
   \,| \,\frac{i}{i \partial_t - H} \,|\, 0, - \mbox{\boldmath $R$} 
   \rangle_\gamma  \cdot {}_\delta \langle z_0,
   \mbox{\boldmath $z$} - \mbox{\boldmath $R$}
   \,| \,\frac{i}{i \partial_t - H} \,| \,- 
   \frac{T}{2}, \mbox{\boldmath $y$} \rangle_{g_1} \ \ \ \ \nonumber \\
   &+& \!\! {}_{f_1} \langle \frac{T}{2}, 
   \mbox{\boldmath $x$} \,| \,
   \frac{i}{i \partial_t - H} \,| \,
   0, - \mbox{\boldmath $R$} \rangle_{\gamma}
   \cdot {}_{\delta} \langle z_0, \mbox{\boldmath $z$} - 
   \mbox{\boldmath $R$} \,| \,
   \frac{i}{i \partial_t - H} \,| \,z_0^\prime, 
   \mbox{\boldmath $z$}^\prime \rangle_{\alpha} \cdot {}_{\beta}
   \langle z_0^\prime, 
   \mbox{\boldmath $z$}^\prime \,| \,\frac{i}{i \partial_t -H} | 
   - \frac{T}{2},
   \mbox{\boldmath $y$}\rangle_{g_1} \ \ \ \ \nonumber \\
   &-& \!\! {}_{f_1} \langle \frac{T}{2}, \mbox{\boldmath $x$}
   \,|\, \frac{i}{i \partial_t - H} \,| \,
   - \frac{T}{2}, \mbox{\boldmath $y$} \rangle_{g_1} \cdot
   {}_{\delta} \langle z_0, \mbox{\boldmath $z$} - 
   \mbox{\boldmath $R$} \,| 
   \,\frac{i}{i \partial_t - H}
   \, | \,  z_0^\prime, \mbox{\boldmath $z$}^\prime \rangle_{\alpha}
   \cdot {}_{\beta} \langle z_0^\prime, 
   \mbox{\boldmath $z$}^\prime \,| \, 
   \frac{i}{i \partial_t - H} \,| \, 0, - \mbox{\boldmath $R$} 
   \rangle_{\gamma} \,\Bigr] \ \ \ \ \nonumber \\
   &+& i \,z_0 \,\frac{1}{2} \,
   {\{ \Omega , \tilde{O}_a \}}_{\gamma \delta} \nonumber \\
   &\times& \!\!
   \Bigl[ \,{}_{f_1}
   \langle\frac{T}{2}, \mbox{\boldmath $x$} \,| \, 
   \frac{i}{i \partial_t - H} \,| \,
   0, -\mbox{\boldmath $R$} \rangle_{\gamma} \cdot
   {}_ {\delta} \langle z_0, 
   \mbox{\boldmath $z$} - \mbox{\boldmath $R$}
   \,| \,\frac{i}{i \partial_t - H} \,|\, - 
   \frac{T}{2}, \mbox{\boldmath $y$} \rangle {}_{g_1} \nonumber \\
   &-& \!\! {}_{\delta} \langle z_0, 
   \mbox{\boldmath $z$} - \mbox{\boldmath $R$} \,| \,
   \frac{i}{i \partial_t - H} \,| \,
   0, - \mbox{\boldmath $R$} \rangle_{\gamma}
   \cdot {}_{f_1} \langle \frac{T}{2}, \mbox{\boldmath $x$} \,| \,
   \frac{i}{i \partial_t - H} \,| \,- \frac{T}{2}, 
   \mbox{\boldmath $y$} \rangle {}_{g_1} \,\Bigr] \,\Biggr\} 
   \nonumber \\
   &\times& \prod^{N_c}_{j = 2} \,\,\,\bigl[\, {}_{f_j}
   \langle \frac{T}{2}, \mbox{\boldmath $x$}
   \,| \,\frac{i}{i \partial_t - H} \,| \,
   - \frac{T}{2}, \mbox{\boldmath $y$} \rangle_{g_j} \,\bigr] \,
   \cdot \,
   \,\,\,\exp \,[\,N_c \,\mbox{Sp} \log (\,i \partial_t - H)
   \,+ \,i \,\frac{I}{2} 
   \,\int \Omega_a^2 \,d t \,\,]  \,\, , \label{omega1}
\end{eqnarray}
and
\begin{eqnarray}
   &\,& \langle N (\mbox{\boldmath $P$}) \,| \,\psi^\dagger (0) \,
   O_a \,\psi (z) \,| \,
   N (\mbox{\boldmath $P$}) \rangle^{\Delta m_s}  \nonumber \\
   &=& \frac{1}{Z} \,\,{\,\tilde{\Gamma}}^{\{ f \}} \,\,
   {\tilde{\Gamma}}^{{\{ g \}}^\dagger} \,\,N_c \,\, 
   \int \,\,d^3 x \,\,d^3 y  \,\,e^{\,-i \mbox{\boldmath $P$} 
   \cdot \mbox{\boldmath $x$}} 
   \,\,e^{\,i \mbox{\boldmath $P$} \cdot \mbox{\boldmath $y$}}
   \,\,\int \,\,d^3 R \,\,\int \,\,{\cal D} {\cal A} \nonumber \\
   &\times& \! \Biggl\{ \,
   \int d^3 z^{\prime} \,\,d z_0^\prime \,\,\,\,
   i \, \Delta H_{\alpha \beta} (z_0^\prime) \,\,
   {(A^\dagger (0) O_a A(z_0))}_{\gamma \delta} \nonumber \\
   &\times& \!\!
   \Bigl[ \,{}_{f_1}
   \langle\frac{T}{2}, \mbox{\boldmath $x$} \,| \, 
   \frac{i}{i \partial_t - H} \,| \,
   z_0^\prime, \mbox{\boldmath $z$}^\prime \rangle_{\alpha} \cdot
   {}_ {\beta} \langle z_0^\prime, \mbox{\boldmath $z$}^\prime 
   \,| \,\frac{i}{i \partial_t - H} \,|\, 0, - \mbox{\boldmath $R$} 
   \rangle_\gamma  \cdot {}_\delta \langle z_0,
   \mbox{\boldmath $z$} - \mbox{\boldmath $R$}
   \,| \,\frac{i}{i \partial_t - H} \,| \,- 
   \frac{T}{2}, \mbox{\boldmath $y$} \rangle_{g_1} \ \ \ \ \nonumber \\
   &+& \!\! {}_{f_1} \langle \frac{T}{2}, 
   \mbox{\boldmath $x$} \,| \,
   \frac{i}{i \partial_t - H} \,| \,
   0, - \mbox{\boldmath $R$} \rangle_{\gamma}
   \cdot {}_{\delta} \langle z_0, \mbox{\boldmath $z$} - 
   \mbox{\boldmath $R$} \,| \,
   \frac{i}{i \partial_t - H} \,| \,z_0^\prime, 
   \mbox{\boldmath $z$}^\prime \rangle_{\alpha} \cdot {}_{\beta}
   \langle z_0^\prime, 
   \mbox{\boldmath $z$}^\prime \,| \,\frac{i}{i \partial_t -H} | 
   - \frac{T}{2},
   \mbox{\boldmath $y$}\rangle_{g_1} \ \ \ \ \nonumber \\
   &-& \!\! {}_{f_1} \langle \frac{T}{2}, \mbox{\boldmath $x$}
   \,|\, \frac{i}{i \partial_t - H} \,| \,
   - \frac{T}{2}, \mbox{\boldmath $y$} \rangle_{g_1} \cdot
   {}_{\delta} \langle z_0, \mbox{\boldmath $z$} - 
   \mbox{\boldmath $R$} \,| 
   \,\frac{i}{i \partial_t - H}
   \, | \,  z_0^\prime, \mbox{\boldmath $z$}^\prime \rangle_{\alpha}
   \cdot {}_{\beta} \langle z_0^\prime, 
   \mbox{\boldmath $z$}^\prime \,| \, 
   \frac{i}{i \partial_t - H} \,| \, 0, - \mbox{\boldmath $R$} 
   \rangle_{\gamma} \,\Bigr] \,\Biggr\} 
   \nonumber \\
   &\times& \prod^{N_c}_{j = 2} \,\,\,\bigl[\, {}_{f_j}
   \langle \frac{T}{2}, \mbox{\boldmath $x$}
   \,| \,\frac{i}{i \partial_t - H} \,| \,
   - \frac{T}{2}, \mbox{\boldmath $y$} \rangle_{g_j} \,\bigr] \,
   \cdot \,
   \,\,\,\exp \,[\,N_c \,\mbox{Sp} \log (\,i \partial_t - H)
   \,+ \,i \,\frac{I}{2} 
   \,\int \Omega_a^2 \,d t \,\,]  \,\, . \label{delms}
\end{eqnarray}
We shall treat these three contributions to the PDF in separate
subsections below.

\subsection{$O (\Omega^0)$ contributions to PDF}

Although we do not need any essential change for the derivation
of the $O (\Omega^0)$ contribution, 
we recall here some main ingredients, since it is
useful for understanding the following manipulation.
We first introduce the eigenstates $| m \rangle$ and the
associated eigenenergies $E_m$ of the static Dirac Hamiltonian $H$,
satisfying
\begin{equation}
 H \vert m \rangle = E_m \vert m \rangle .
\end{equation}
The spectral representation of the single quark Green function is 
then given as
\begin{eqnarray}
   {}_{\alpha} \langle\mbox{\boldmath $x$}, t \,|\,
   \frac{i}{i \partial_t - H} \,| \,
   \mbox{\boldmath $x$}^\prime, t^\prime\rangle_{\beta}
   &=& \theta(t - t^\prime) \,\,\sum_{m>0} \,\,
   e^{\,- i E_m (t - t^\prime)} \,\,
   {}_{\alpha} \langle\mbox{\boldmath $x$}\,| \,m \rangle 
   \langle m \,| \,\mbox{\boldmath $x$}^\prime \rangle_{\beta}
   \nonumber \\
   &-& \theta(t^\prime - t) \,\,\sum_{m<0} \,\, 
   e^{\,-i E_m (t - t^\prime)} \,\,
   {}_{\alpha} \langle\mbox{\boldmath $x$} \,| \,m \rangle
   \langle m \,| \,
   \mbox{\boldmath $x$}^\prime\rangle_{\beta} \,\, . \ \
\end{eqnarray}
Using this equation together with the identity
\begin{equation}
   \langle \mbox{\boldmath $z$} - \mbox{\boldmath $R$} \, | \, 
   \ \ = \ \ \langle -\mbox{\boldmath $R$} \,| \,\,
   e^{\,i \mbox{\boldmath $p$} \cdot \mbox{\boldmath $z$}} \,\, ,
\end{equation}
with $\mbox{\boldmath $p$}$ being the momentum operator,
we can perform the 
integration over $\mbox{\boldmath $R$}$ in (\ref{base2}). 
The resultant expression is then put into (\ref{base1}) to carry out
the integration over $z_0$. This leads to the following
expression for the quark distribution function :
\begin{equation}
   q (x ; \Omega^0) \ \ = \ \ \int \,
   \Psi_{Y T T_3 ; J J_3}^{(n)^*} [\xi_A] \,\,\,O^{(0)} [\xi_A] \,\,
   \Psi_{Y T T_3 ; J J_3}^{(n)} [\xi_A] \,\,d \xi_A . \label{matelm}
\end{equation}
Here $O^{(0)} [ \xi_A ]$ is an $O(\Omega^0)$ effective operator 
given by
\begin{equation}
   O^{(0)} [\xi_A] \ \ = \ \ M_N \,\frac{N_c}{2} \,
   \Bigl( \,\sum_{n \leq 0} - \sum_{n > 0} \,\Bigr) \,
   \langle n | \,\tilde{O}_a \delta (x M_N - E_n - p_3) \,
   | n \rangle .
\end{equation}
Note that it is still a functional of the collective coordinates 
$\xi_A$ that specify the orientation of the hedgehog soliton in 
the collective coordinate space.
The physical baryons are identified as rotational states of this 
collective motion and the corresponding wave functions are denoted 
as $\Psi^{(n)}_{Y T T_3 ; J J_3} [\xi_A ]$, which belongs to a 
SU(3) representation of dimension $n$ with relevant spin-flavor 
quantum numbers. Using the standard Wigner rotation matrix 
(or D-function) of SU(3) group, they are represented as 
\begin{equation}
 \Psi_{Y T T_3 ; J J_3}^{(n)} [\xi_a] = {(-1)}^{J + J_3}
 \sqrt{n} \,D_{\mu, \nu}^{(n)} (\xi_a)
\end{equation}
with $\mu = (Y T T_3)$ and $\nu = (Y^\prime = 1, J J_3)$.
In the present study, we are interested in the quark distribution 
functions in the nucleon, so that we can set $Y = 1$ and 
$T = J = 1/2$.

The general formula can now be used to derive some more explicit form
of the $O(\Omega^0)$ contribution to the quark distribution functions.
We first consider the unpolarized distributions.
For the flavor singlet case, we take 
\begin{equation}
 \tilde{O}_{a=0} = A^\dagger \lambda_0 A \,
 (1 + \gamma^0 \gamma^3) = 1 + \gamma^0 \gamma^3 ,
\end{equation}
so that we find that
\begin{equation}
 O^{(0)} [\xi_A] = M_N \frac{N_c}{2} \left(
 \sum_{n \leq 0} - \sum_{n > 0} \right) \,
 \langle n \vert (1 + \gamma^0 \gamma^3) \delta_n \vert n \rangle ,
\end{equation}
with the abbreviation $\delta_n = \delta (x M_N - E_n - p_3)$.
This then gives
\begin{equation}
 q^{(0)} (x ; \Omega^0) = {\langle 1 \rangle}_p \cdot f(x) ,
\end{equation}
with the definition
\begin{equation}
 f(x) = M_N \frac{N_c}{2} 
 \left( \sum_{n \leq 0} - \sum_{n > 0}  \right) \,
 \langle n \vert (1 + \gamma^0 \gamma^3) \delta_n
 \vert n \rangle .
\end{equation}
Here and hereafter, ${\langle O \rangle}_B$ should be understood
as an abbreviated notation of the matrix element of a collective
operator $O$ between a baryon state $B$ (mostly, the spin-up
proton state) with appropriate quantum numbers, i.e.
\begin{equation}
 {\langle O \rangle}_B \equiv \int
 \Psi_{Y T T_3 ; J J_3}^{(n)*} [\xi_A] \,O [\xi_A] \,
 \Psi_{Y T T_3 ; J J_3}^{(n)} [\xi_A] \,d \xi_A .
\end{equation}
In  the flavor nonsinglet case $(a = 3 \ \mbox{or} \ 8)$, 

\begin{equation}
 \tilde{O}_a = A^\dagger \lambda_a A \,
 (1 + \gamma^0 \gamma^3) = D_{ab} \lambda_b \,
 (1 + \gamma^0 \gamma^3) ,
\end{equation}
we have
\begin{eqnarray}
 O^{(a)} [\xi_A] &=& D_{ab} \cdot M_N \frac{N_c}{2}
 \left( \sum_{n \leq 0} - \sum_{n > 0} \right)
 \langle n \vert \lambda_a (1 + \gamma^0 \gamma^3) \delta_n
 \vert n \rangle \nonumber \\
 &=& D_{a8} \cdot M_N \frac{N_c}{2}
 \left( \sum_{n \leq 0} - \sum_{n > 0} \right)
 \langle n \vert \lambda_8 (1 + \gamma^0 \gamma^3) \delta_n
 \vert n \rangle \nonumber \\
 &=& \frac{D_{a8}}{\sqrt{3}} \cdot M_N \frac{N_c}{2}
 \left( \sum_{n \leq 0} - \sum_{n > 0} \right)
 \langle n \vert (1 + \gamma^0 \gamma^3) \delta_n
 \vert n \rangle .
\end{eqnarray}
Here, we have used the generalized hedgehog symmetry of the classical
configuration (\ref{hedgehog}). This then gives, for
$a = 3 \ \mbox{or} \ 8$,
\begin{equation}
 q^{(a)} (x ; \Omega^0) = 
 \langle \frac{D_{a8}}{\sqrt{3}} \rangle_p
 \cdot f(x) .
\end{equation}
Turning to the longitudinally polarized distribution, we take
\begin{equation}
 \tilde{O}_a = A^\dagger \lambda_0 A \,
 (1 + \gamma^0 \gamma^3) \gamma_5 = \gamma_5 + \Sigma_3 ,
\end{equation}
for the flavor-singlet case, so that we find
\begin{equation}
 \Delta q^{(0)} (x ; \Omega^0) = 0 .
\end{equation}
On the other hand, for the flavor non-singlet case we obtain
\begin{equation}
 \tilde{O}_a = A^\dagger \lambda_a A \,
 (1 + \gamma^0 \gamma^3) \gamma_5 = 
 D_{ab} \lambda_b \,(\gamma_5 + \Sigma_3) .
\end{equation}
This gives
\begin{eqnarray}
 O^{(a)} [\xi_A] &=& M_N \frac{N_c}{2}
 \left( \sum_{n \leq 0} - \sum_{n > 0} \right)
 \langle n \vert D_{ab} \lambda_b (\gamma_5 + \Sigma_3) \delta_n
 \vert n \rangle \nonumber \\
 &=& D_{a3} \cdot M_N \frac{N_c}{3} 
 \left( \sum_{n \leq 0} - \sum_{n > 0} \right)
 \langle n \vert \lambda_3 (\gamma_5 + \Sigma_3) \delta_n
 \vert n \rangle .
\end{eqnarray}
We therefore have, for $a = 3 \ \mbox{or} \ 8$,
\begin{equation}
 \Delta q^{(a)} (x ; \Omega^0) = 
 {\langle - \,D_{a3} \rangle}_{p \uparrow} \cdot g(x) ,
\end{equation}
with
\begin{equation}
 g(x) = - M_N \frac{N_c}{2}
 \left( \sum_{n \leq 0} - \sum_{n > 0} \right)
 \langle n \vert \lambda_3 (\gamma_5 + \Sigma_3) \delta_n
 \vert n \rangle .
\end{equation}

\subsection{$O(\Omega^1)$ contribution to PDF}

There is some controversy in the treatment of the $O(\Omega^1)$ 
term in the CQSM. The dispute began after our finding of the novel 
$1 / N_c$ correction (or more explicitly the first order rotational
correction in the collective angular velocity $\Omega$)
to some isovector observables like the isovector 
part of the nucleon axial-vector coupling constant $g_A^{(3)}$ or 
the isovector magnetic moment $\mu_{I = 1}$ \cite{WW93}.
We showed that this new $1 / N_c$ correction, which is entirely
missing in the theoretical framework of the intimately-connected
effective meson theory, i.e. the Skyrme model, 
plays just a desirable role in solving the 
long-standing $g_A$ problem inherent in the soliton model based on 
the hedgehog configuration \cite{WW93},\cite{CBGPPWW94}.
According to Schechter and Weigel \cite{SW95A},\cite{SW95B}, 
however, this $O (\Omega^1)$ contribution originates from the 
ordering ambiguity of the collective operators and it breaks the 
G-parity symmetry of strong interactions. We agree that the operator 
ordering ambiguity is unavoidable when going from a classical theory 
to a quantum theory. Different choice of ordering would in general
define different quantum theory.
It was shown, however, that the existence 
of this new $O(\Omega^1)$ contribution is a natural consequence of
a physically reasonable choice of operator ordering that keeps the 
time-order of the relevant operators and that this 
$O(\Omega^1)$ contribution to $g_A^{(3)}$ is nothing 
incompatible with any symmetry of strong interactions including the
G-parity symmetry \cite{W95}\nocite{W96}--\cite{CGP95}.
We also recall the fact that this
time-order-keeping quantization procedure is nothing extraordinary
in that it gives the same answer as the so-called cranking approach
familiar in the nuclear many-body theory \cite{W96}.
(Alkofer and Weigel also claimed that the new $O(\Omega^1)$ term
breaks the celebrated PCAC relation \cite{AW93}.
Here we do not argue this problem further,
since our attitude is that this problem does not exist either
within the framework of the SU(2) CQSM as discussed
in Ref.~\cite{W96}.)
Summarizing our understanding about this problem up to this point,
the ordering ambiguity of the collective operator in principle
exists, but a physically reasonable time-order-keeping quantization
procedure leads to the desired $O(\Omega^1)$ contribution to
$g_A^{(3)}$, while causing no problem at least in the flavor SU(2)
version of the CQSM.
However, Prasza{\l}owicz et al. noticed an unpleasant feature of the 
time-order-keeping quantization procedure in the flavor SU(3) version 
of the CQSM \cite{PWG99}.
That is, it leads to nondiagonal elements in the moment
of inertia tensor of the soliton, which may destroy the basic 
theoretical framework of the soliton model.
Since there is no such problem in the SU(2) CQSM, the cause of this
trouble seems to be attributed to the incompatibility of the
time-order-keeping quantization
procedure with the basic dynamical assumption of the SU(3) CQSM,
i.e. the so-called trivial embedding of the SU(2) soliton
configuration followed by the SU(3) symmetric collective quantization.
In the absence of satisfactory resolution to this problem, they 
advocated to use phenomenologically favorable procedure, which 
amounts to dropping some theoretically contradictory terms by hand.
In the present study, we shall basically follow this procedure.
As we shall discuss below, however, the operator ordering problem is
even more complicated in our study 
of quark distribution functions, since we must handle here quark 
bilinear operators which are nonlocal also in time coordinates.

In our formulation of the $O(\Omega^1)$ contribution to 
the distribution function, the ordering problem arises when
handling the product of operators
\begin{equation}
 \Omega_{\alpha \beta} (z_0^\prime) \,{\left(
 A^\dagger (0) O_a A(z_0) \right)}_{\gamma \delta} ,
\end{equation}
in (\ref{omega1}). In the previous paper, we adopted the ordering
\begin{eqnarray}
   \Omega_{\alpha \beta} (z'_0) \,
   {(A^\dagger (0) O_a A(z_0))}_{\gamma \delta} 
   &\longrightarrow&
   [ \theta (z'_0, 0, z_0) + \theta (z'_0, z_0, 0) ]
   \,\Omega_{\alpha \beta} \,\tilde{O}_{\gamma \delta} \nonumber \\
   &+& [ \theta (0, z_0, z'_0) + \theta (z_0, 0, z'_0) ] \,
   \tilde{O}_{\gamma \delta} \Omega_{\alpha \beta}  \nonumber \\
   &+& \ \theta(0, z'_0, z_0) \,{(O_a)}_{\gamma^\prime \delta^\prime}
   \,A^\dagger_{\gamma \gamma^\prime} \,
   \Omega_{\alpha \beta} \,A_{\delta^\prime \delta} \nonumber \\
   &+& \ \theta(z_0, z'_0, 0) \,
   {(O_a)}_{\gamma^\prime \delta^\prime} \,
   A_{\delta^\prime \delta} \,\Omega_{\alpha \beta} \,
   A^\dagger_{\gamma \gamma^\prime} ,
\end{eqnarray}
because it is a procedure faithful to the time-order of all the
relevant collective operators. In consideration of the existence of
operator ordering ambiguity in quantization, we use here somewhat 
simpler ordering procedure specified as
\begin{eqnarray}
   \Omega_{\alpha \beta} (z'_0) \,
   {(A^\dagger (0) O_a A(z_0))}_{\gamma \delta}
   &\longrightarrow& [ \theta (z'_0, 0, z_0) + \theta (z'_0, z_0, 0) ]
   \,\Omega_{\alpha \beta} \,\tilde{O}_{\gamma \delta} \nonumber \\
   &+& [ \theta (0, z_0, z'_0) + \theta (z_0, 0, z'_0) ] \,
   \tilde{O}_{\gamma \delta} \Omega_{\alpha \beta}  \nonumber \\
   &+& \ \theta(0, z'_0, z_0) \,
   \frac{1}{2} \,\{ \Omega_{\alpha \beta} ,\tilde{O}_{\gamma \delta}
   \} \nonumber \\
   &+& \ \theta(z_0, z'_0, 0) \,
   \frac{1}{2} \,\{ \Omega_{\alpha \beta} ,\tilde{O}_{\gamma \delta}
   \} . \label{ordering}
\end{eqnarray}
The difference between the new and the old quantization procedures 
turns out to be that $O^{(1)}_{B^\prime}$ term in eq.(67)
of Ref.\cite{WK99} is absent in the new procedure.
The operator ordering ambiguity occurs also for the quantity
$\frac{1}{2} \{ \Omega, \tilde{O}_a\}_{\gamma \delta}$
in (\ref{omega1}), which corresponds to the first order rotational
correction arising from the nonlocality (in time) of the operator
$A^{+} (0) O_a A (z_0)$.
To explain it, we first recall the quantization rule of the SU(3) 
collective rotation given as
\begin{equation}
 \Omega = \frac{1}{2} \Omega_a \lambda_a ,
\end{equation}
with
\begin{equation}
 J_a \equiv - R_a = \left\{ \begin{array}{ll}
 I_1 \Omega_a - \frac{2}{\sqrt{3}} \Delta m_s K_1 D_{8a} &
 \ \ \ (a = 1,2,3) , \\
 I_2 \Omega_a - \frac{2}{\sqrt{3}} \Delta m_s K_2 D_{8a} &
 \ \ \ (a = 4,5,6,7) , \\
 \hspace{10mm} \sqrt{3} / 2 & \ \ \ (a = 8).
 \end{array} \right. \, \label{qrulems}
\end{equation}
Here $R_a$ is the right rotation generator also familiar in the SU(3)
Skyrme model. Note that only $a = 1,2,3$ component of $J_a = -R_a$
can be interpreted as the standard angular momentum operators.
In the above equations, $I_1,I_2$ and $K_1,K_2$ are the components
of the moment-of-inertia tensor of the soliton defined by
\begin{eqnarray}
 I_{ab} &=& \frac{N_c}{2} \sum_{m \geq 0, n < 0}
 \frac{\langle n \vert \lambda_a \vert m \rangle
 \langle m \vert \lambda_b \vert n \rangle}{E_m - E_n} , \\ \label{momi}
 K_{ab} &=& \frac{N_c}{2} \sum_{m \geq 0, n < 0}
 \frac{\langle n \vert \lambda_a \vert m \rangle
 \langle m \vert \lambda_b \gamma^0 \vert n \rangle}{E_m - E_n} ,
\end{eqnarray}
which reduce to the form
\begin{eqnarray}
 I_{ab} &=& \mbox{diag} 
 \left( I_1,I_1,I_1,I_2,I_2,I_2,I_2,0 \right) , \\
 K_{ab} &=& \mbox{diag} 
 \left( K_1,K_1,K_1,K_2,K_2,K_2,K_2,0 \right) , \label{momk}
\end{eqnarray}
because of the hedgehog symmetry.
Setting $\Delta m_s = 0$, for the moment, to keep the discussion
below simpler, we therefore obtain
\begin{equation}
 \left\{ \tilde{O}_a , \Omega \right\} = 
 \frac{1}{2 I_1} \left\{ D_{ab} \lambda_b \bar{O} , 
 J_i \lambda_i \right\} + 
 \frac{1}{2 I_2} \left\{ D_{ab} \lambda_b \bar{O} , 
 J_K \lambda_K \right\} ,
\end{equation}
where the summation over the repeated indices is understood with
$i$ running from $1$ to $3$, and with $K$ from $4$ to $7$. To keep
compliance with the new operator ordering procedure (\ref{ordering})
explained above, we assume the symmetrization of the operator products
as
\begin{eqnarray}
 D_{ab} J_c &\longrightarrow& 
 \frac{1}{2} \left\{
 D_{ab} , J_c \right\} \\
 J_c D_{ab} &\longrightarrow& 
 \frac{1}{2} \left\{
 D_{ab}, J_c \right\} ,
\end{eqnarray}
prior to quantization. This amounts to the replacement
\begin{equation}
 \left\{ \tilde{O}_a , \Omega \right\} \longrightarrow
 {\left\{ \tilde{O}_a , \Omega \right\}}^S ,
\end{equation}
with
\begin{equation}
 {\left\{ \tilde{O}_a , \Omega \right\}}^S = 
 \frac{1}{2 I_1} \left\{ D_{ab}, J_i \right\}
 \left\{ \lambda_b, \lambda_i \right\} + 
 \frac{1}{2 I_2} \left\{ D_{ab}, J_K \right\}
 \left\{ \lambda_b, \lambda_K \right\} .
\end{equation}
Now collecting all the terms, which is first order in $\Omega$,
we arrive at the following expression for the $O(\Omega^1)$
effective operator to be sandwiched between the rotational wave
functions as in (\ref{matelm}). It is given by 
\begin{equation}
 O^{(1)} [\xi_A] = O_A^{(1)} + O_B^{(1)} + O_C^{(1)} ,
\end{equation}
where
\begin{eqnarray}
 O_A^{(1)} &=& M_N \frac{N_c}{2} \sum_{m > 0,n \leq 0}
 \frac{1}{E_m - E_n} \nonumber \\
 &\,& \times \left[
 \langle n \vert \tilde{O}_a (\delta_n + \delta_m) \vert m \rangle
 \langle m \vert \Omega \vert n \rangle + 
 \langle n \vert \Omega \vert m \rangle
 \langle m \vert \tilde{O}_a (\delta_n + \delta_m) 
 \vert n \rangle \right] ,\\
 O_B^{(1)} &=& M_N \frac{N_c}{2} 
 (\sum_{m \leq 0,n \leq 0} - \sum_{n > 0, m > 0})
 \frac{1}{E_m - E_n} \nonumber \\
 &\,& \times \left[
 \langle n \vert \tilde{O}_a (\delta_n - \delta_m) \vert m \rangle
 \langle m \vert \Omega \vert n \rangle + 
 \langle n \vert \Omega \vert m \rangle
 \langle m \vert \tilde{O}_a (\delta_n - \delta_m) 
 \vert n \rangle \right] ,
\end{eqnarray}
while
\begin{eqnarray}
 O_C^{(1)} &=& \frac{1}{2 I_1} \,J_i \,
 \frac{N_c}{2} \frac{d}{dx} 
 \left(\sum_{n \leq 0} - \sum_{n > 0} \right)
 \langle n \vert \lambda_i \,
 \bar{O} \delta_n \vert n \rangle \nonumber \\
 &+& \frac{1}{2 I_2} \,J_K \,
 \frac{N_c}{2} \frac{d}{dx} 
 \left(\sum_{n \leq 0} - \sum_{n > 0} \right)
 \langle n \vert \lambda_K \,
 \bar{O} \delta_n \vert n \rangle .
\end{eqnarray}
for the flavor singlet case, and
\begin{eqnarray}
 O_C^{(1)} &=& \frac{1}{4 I_1} \left\{ D_{ab}, J_i \right\}
 \frac{N_c}{2} \frac{d}{dx} 
 \left(\sum_{n \leq 0} - \sum_{n > 0} \right)
 \langle n \vert \left\{ \lambda_b, \lambda_i \right\}
 \bar{O} \delta_n \vert n \rangle \nonumber \\
 &+& \frac{1}{4 I_2} \left\{ D_{ab}, J_K \right\}
 \frac{N_c}{2} \frac{d}{dx} 
 \left(\sum_{n \leq 0} - \sum_{n > 0} \right)
 \langle n \vert \left\{ \lambda_b, \lambda_K \right\}
 \bar{O} \delta_n \vert n \rangle .
\end{eqnarray}
for the flavor nonsinglet case.
As was done in Ref.~\cite{WK99}, it is convenient to treat
$O_A^{(1)}$ and $O_B^{(1)}$ in a combined way, i.e.
in such a way that it is given 
as a sum of two parts, respectively containing symmetric and 
antisymmetric pieces with respect to the collective space operators
$D_{a b}$ and $J_c$ as
\begin{equation}
 O_A^{(1)} + O_B^{(1)} = O_{\{A,B\}}^{(1)} + O_{[A,B]}^{(1)} .
\end{equation}
For obtaining the explicit forms of $O_{ \{ A,B \}}^{(1)}$ and 
$O_{ [ A.B ]}^{(1)}$, we will treat the two cases separately.
First is the case in which $O_a$ is a flavor singlet operator
as $O_a = \bar{O}$.
In this case, we have
\begin{eqnarray}
 O_{\{A,B\}}^{(1)} &=& - M_N \frac{N_c}{4 I_1} J_i
 \left( \sum_{m = all,n \leq 0} - \sum_{m = all, n > 0} \right)
 \frac{1}{E_m - E_n} \nonumber \\
 &\,& \hspace{10mm} \times 
 \left[ \langle n \vert \bar{O} \delta_n \vert m \rangle
 \langle m \vert \lambda_i \vert n \rangle + 
 \langle n \vert \lambda_i \vert m \rangle
 \langle m \vert \bar{O} \delta_n \vert n \rangle \right] , \\
 O_{[A,B]}^{(1)} &=& \hspace{10mm} 0 \,\, .
\end{eqnarray}
On the other hand, if $O_a$ is a flavor nonsinglet operator such 
as $O_a = \lambda_a \bar{O}$, we find
\begin{eqnarray}
 O_{\{A,B\}}^{(1)} 
 &=& - M_N \frac{N_c}{4 I_1} \frac{1}{2}
 \left\{ D_{ab}, J_i \right\}
 \left( \sum_{m = all,n \leq 0} - \sum_{m = all, n > 0} \right)
 \frac{1}{E_m - E_n} \nonumber \\
 &\,& \hspace{10mm} \times \left[ \langle n \vert \lambda_b \bar{O} 
 \delta_n \vert m \rangle \langle m \vert \lambda_i \vert n \rangle + 
 \langle n \vert \lambda_i \vert m \rangle
 \langle m \vert \lambda_b \bar{O} \delta_n \vert n \rangle \right]
 \nonumber \\
 &\,& - M_N \frac{N_c}{4 I_2} \frac{1}{2}
 \left\{ D_{ab}, J_K \right\}
 \left( \sum_{m = all,n \leq 0} - \sum_{m = all, n > 0} \right)
 \frac{1}{E_m - E_n} \nonumber \\
 &\,& \hspace{10mm} \times \left[ \langle n \vert \lambda_b \bar{O} 
 \delta_n \vert m \rangle \langle m \vert \lambda_K \vert n \rangle + 
 \langle n \vert \lambda_K \vert m \rangle
 \langle m \vert \lambda_b \bar{O} \delta_n \vert n \rangle \right] ,
\end{eqnarray}
and
\begin{eqnarray}
 O_{[A,B]}^{(1)}
 &=& - M_N \frac{N_c}{4 I_1} \frac{1}{2}
 \left[ D_{ab}, J_i \right]
 \left( \sum_{m > 0,n \leq 0} + \sum_{m \leq 0, n > 0} \right)
 \frac{1}{E_m - E_n} \nonumber \\
 &\,& \hspace{10mm} \times \left[ \langle n \vert \lambda_b \bar{O} 
 \delta_n \vert m \rangle \langle m \vert \lambda_i \vert n \rangle - 
 \langle n \vert \lambda_i \vert m \rangle
 \langle m \vert \lambda_b \bar{O} \delta_n \vert n \rangle \right]
 \nonumber \\
 &\,& - M_N \frac{N_c}{4 I_2} \frac{1}{2}
 \left[ D_{ab}, J_K \right]
 \left( \sum_{m > 0,n \leq 0} - \sum_{m \leq 0, n > 0} \right)
 \frac{1}{E_m - E_n} \nonumber \\
 &\,& \hspace{10mm} \times \left[ \langle n \vert \lambda_b \bar{O} 
 \delta_n \vert m \rangle \langle m \vert \lambda_K \vert n \rangle - 
 \langle n \vert \lambda_K \vert m \rangle
 \langle m \vert \lambda_b \bar{O} \delta_n \vert n \rangle \right] ,
 \label{antisym}
\end{eqnarray}
We point out that these expressions also are not completely free 
from operator-ordering ambiguities. If we symmetrize the order of 
two operators $\Omega_{\alpha \beta}$ and $\tilde{O}_{\gamma \delta}$
in the first and the second term of eq.(\ref{antisym}),
the antisymmetric term $O_{[ A,B ]}^{(1)}$ does not appear from the
first.
A favorable aspect of this symmetrization procedure is that it does 
not cause an internal inconsistency of the SU(3) CQSM, which was 
first pointed out by Prasza{\l}owicz et al.~\cite{PWG99}
Unfortunately, however,
it also eliminates the phenomenologically-welcome first-order 
rotational correction to $g_A^{(3)}$, the sprout of which is contained 
in the first term of (\ref{antisym}). As repeatedly emphasized, the
presence of this novel $1 / N_c$ correction itself is nothing 
incompatible with any symmetry of strong interaction. Although
it is not a completely satisfactory procedure, we therefore retain
only the first term of (\ref{antisym}) and abandon the second one,
which precisely corresponds to the symmetry-preserving approach
advocated by Prasza{\l}owicz et al. 

Now we consider the concrete case again. For the flavor singlet 
unpolarized distribution, we find there exists no $O(\Omega^1)$ 
contribution, i.e.
\begin{equation}
 q^{(0)} (x, \Omega^1) = 0 .
\end{equation}
In the flavor nonsinglet case, the $O(\Omega^1)$ contribution 
consists of two terms as 
\begin{equation}
 O^{(1)} [ \xi_A ] = O^{(1)}_{ \{A,B \} } + O^{(1)}_C .
\end{equation}
Here
\begin{eqnarray}
 O^{(1)}_{ \{A,B \} } &=& \frac{1}{2} \{ D_{a b}. R_i \} \cdot M_N
 \frac{N_c}{4 I_1} \left( \sum_{m = all, n \leq 0}
 - \sum_{m = all, n > 0} \right) \frac{1}{E_m - E_n} \nonumber \\
 &\,& \times
 [ \langle n | \lambda_b (1 + \gamma^0 \gamma^3) \delta_n | m \rangle
 \langle m | \lambda_i | n \rangle + 
 \langle n | \lambda_i | m \rangle
 \langle m | \lambda_b (1 + \gamma^0 \gamma^3 ) \delta_n | n \rangle ]
 \nonumber \\
 &+& \frac{1}{2} \{ D_{a b}. R_K \} \cdot M_N 
 \frac{N_c}{4 I_2} \left( \sum_{m = all, n \leq 0} 
 - \sum_{m = all, n > 0} \right) \frac{1}{E_m - E_n} \nonumber \\
 &\,& \times
 [ \langle n | \lambda_b (1 + \gamma^0 \gamma^3) \,
 \delta_n | m \rangle 
 \langle m | \lambda_K | n \rangle + 
 \langle n | \lambda_K | m \rangle
 \langle m | \lambda_b (1 + \gamma^0 \gamma^3) 
 \delta_n | n \rangle \nonumber ] \\
 &=& \,\sum_{i = 1}^3 \{ D_{a i}, R_i \} \cdot 
 \frac{M_N}{I_1} \cdot \frac{1}{3} \sum_{j = 1}^3
 \frac{N_c}{2} \left( \sum_{m = all, n \leq 0}
 - \sum_{m = all, n > 0} \right) \nonumber \\ 
 &\,& \hspace{25mm} \times \,\frac{1}{E_m - E_n} \,
 \langle n | \lambda_j | m \rangle 
 \langle m | \lambda_j  (1 + \gamma^0 \gamma^3) 
 \delta_n | n \rangle \nonumber \\
 &+& \sum_{K = 4}^7 \{ D_{a K}, R_K \} \frac{M_N}{I_2} \cdot
 \frac{N_c}{2} \left( \sum_{m = all, n \leq 0} 
 - \sum_{m = all, n > 0} \right) \nonumber \\
 &\,& \hspace{25mm} \times \,\frac{1}{E_m - E_n} \,
 \langle n | \lambda_4 | m \rangle 
 \langle m | \lambda_4 (1 + \gamma^0 \gamma^3) 
 \delta_n | n \rangle .
\end{eqnarray}
In deriving the last equality, we have made use of the generalized
hedgehog symmetry of the static soliton configuration. The explicit
summation symbol for the repeated indices have been restored here
for clarity. For the second contribution to 
$O^{(1)} [ \xi_A ]$, we have 
\begin{eqnarray}
 O^{(1)}_C &=& - \sum_{i = 1}^3 \{ D_{8 i}, R_i \} \frac{1}{2 I_1}
 \frac{1}{3} \sum_{j = 1}^3 \frac{N_c}{2} \frac{d}{d x}
 \left( \sum_{m = all, n \leq 0} - \sum_{m = all, n > 0} \right)
 \nonumber \\
 &\,& \hspace{20mm} \times \,\frac{1}{E_m - E_n} \,
 \langle n | \lambda_j | m \rangle
 \langle m | \lambda_j (1 + \gamma^0 \gamma^3) 
 \delta_n | n \rangle \nonumber \\
 &\,& - \,\sum_{K = 4}^7 \{ D_{8 K}. R_K \} \frac{1}{2 I_2}
 \frac{N_c}{2} \frac{d}{d x}
 \left(\sum_{m = all, n \leq 0} - \sum_{m = all, n > 0 } \right) 
 \nonumber \\
 &\,& \hspace{20mm} \times \,\frac{1}{E_m - E_n} \,
 \langle n | \lambda_4 | m \rangle
 \langle m | \lambda_4 (1 + \gamma^0 \gamma^3) 
 \delta_n | n \rangle .
\end{eqnarray}
Here, we have used the identities : 
\begin{eqnarray}
 \frac{1}{3} \sum_{j = 1}^3 \sum_{m = all, M(n)}
 \langle n | \lambda_j | m \rangle 
 \langle m | \lambda_j (1 + \gamma^0 \gamma^3)
 \delta_n | n \rangle \ = \ 
 \sum_{M(n)} \langle n | (1 + \gamma^0 \gamma^3) 
 \delta_n | n \rangle ,
\end{eqnarray}
and
\begin{eqnarray}
 \sum_{m = all, M(n)} \langle n | \lambda_4 | m \rangle
 \langle m | \lambda_4 \,
 (1 + \gamma^0 \gamma^3) \delta_n | n \rangle
 \ = \ \frac{1}{2} \sum_{M(n)}
 \langle n | (1 + \gamma^0 \gamma^3) \delta_n | n \rangle .
\end{eqnarray}
Here and hereafter, $\sum_{M(n)}$ stands for the summation over
the third component of the grand spin of the eigenstate $n$. 
The second identity can be proved as follows
\begin{eqnarray}
 &\,& \!\!\! \sum_{m = all, M(n)} \langle n | \lambda_4 | m \rangle
 \langle m | \lambda_4 (1 + \gamma^0 \gamma^3) \delta_n | n \rangle
 \nonumber \\
 &=& \sum_{M(n)}
 \langle n | \lambda^2_4 (1 + \gamma^0 \gamma^3) \delta_n | n \rangle
 \nonumber \\
 &=& \sum_{M(n)} \langle n | \left(\frac{2}{3} - 
 \frac{1}{2 \sqrt{3}} \lambda_8
 + \frac{1}{2} \lambda_3 \right) 
 (1 + \gamma^0 \gamma^3) \delta_n | n \rangle
 \nonumber \\
 &=& \sum_{M(n)} \langle n | \left(\frac{2}{3} - \frac{1}{2 \sqrt{3}} 
 \cdot  \frac{1}{\sqrt{3}} \right) 
 (1 + \gamma^0 \gamma^3) \delta_n | n \rangle
 \nonumber \\
 &=& \frac{1}{2} \sum_{M(n)}
 \langle n | (1 + \gamma^0 \gamma^3) \delta_n | n \rangle ,
\end{eqnarray}
where the generalized hedgehog symmetry is used again. Now combining
$O^{(1)}_{ \{ A,B \} }$ and $O^{(1)}_C$ terms, the $O(\Omega^1)$
contribution to the flavor nonsinglet $(a = 3 \ \mbox{or} \ 8)$
unpolarized distribution function can be expressed as
\begin{eqnarray}
 q^{(a)} (x ; \Omega^1) &=& \langle \sum_{i = 1}^3
 \{ D_{a i}, R_i \} \rangle_p \cdot \,k_1 (x) \nonumber \\
 &+& \langle \sum_{K = 4}^7 \{ D_{a K}, R_K \} \rangle_p 
 \cdot \,k_2 (x) , \label{k1k2}
\end{eqnarray}
with 
\begin{eqnarray}
 k_1 (x) &=& M_N \frac{1}{2 I_1} \frac{N_c}{2} \frac{1}{3}
 \sum_{j = 1}^3 \left( \sum_{m = all, n \leq 0} 
 - \sum_{m = all, n > 0} \right) \nonumber \\
 &\,& \hspace{10mm} \times \,\langle n | \lambda_j | m \rangle
 \langle m | \lambda_j (1 + \gamma^0 \gamma^3) 
 \left( \frac{\delta_n}{E_m - E_n} - \frac{1}{2} \,
 \delta_n^\prime \right)
 | n \rangle ,
\end{eqnarray}
and
\begin{eqnarray}
 k_2 (x) &=& M_N \frac{1}{2 I_2} \frac{N_c}{2}
 \left( \sum_{m + all, n \leq 0} - \sum_{m = all, n > 0} \right)
 \nonumber \\
 &\,& \hspace{10mm} \times \,\langle n | \lambda_4 | m \rangle
 \langle m | \lambda_4 (1 + \gamma^0 \gamma^3)
 \left( \frac{\delta_n}{E_m - E_n} - \frac{1}{2} \,
 \delta_n^\prime \right) 
 | n \rangle .
\end{eqnarray}
Here we have used the notation $\delta_n \equiv \delta 
(x M_N - E_n - p_3)$ and $\delta^\prime_n \equiv \delta^\prime 
(x M_N - E_n - p_3)$.
Turning to the longitudinally polarized distributions, the
$O(\Omega^1)$ contribution to the flavor singlet distribution
consists of two terms as
\begin{equation}
 O^{(1)} [\xi_A ] \ = \ O^{(1)}_{ \{A,B \} } \ + \ O^{(1)}_C ,
\end{equation}
where
\begin{eqnarray}
 O^{(1)}_{ \{A,B \} } &=& - 2 J_3 \cdot M_N \frac{N_c}{4 I_1}
 \left( \sum_{m = all, n \leq 0} - \sum_{m = all, n > 0} \right)
 \frac{1}{E_m - E_n} \nonumber \\
 &\,& \hspace{35mm} \times \,
 \langle n | \lambda_3 | m \rangle 
 \langle m | (\gamma_5 + \Sigma_3) 
 \delta_n | n \rangle ,
\end{eqnarray}
and
\begin{eqnarray}
 O^{(1)}_C &=& 2 J_3 \cdot \frac{d}{d x} \frac{N_c}{8 I_1}
 \left( \sum_{m = all, n \leq 0} - \sum_{m = all, n > 0} \right)
 \nonumber \\
 &\,&  \hspace{25mm} \times \,
 \langle n | \lambda_3 | m \rangle 
 \langle m | (\gamma_5 + \Sigma_3)
 \delta_n | n \rangle .
\end{eqnarray}
Combining the two terms, we have
\begin{equation}
 \Delta q^{(0)} (x : \Omega^1) \ = \ 
 {\langle 2 J_3 \rangle}_{p \uparrow} \cdot \,e(x) ,
\end{equation}
with
\begin{eqnarray}
 e(x) &=& M_N \frac{N_c}{4 I_1} 
 \left( \sum_{m = all, n \leq 0} - \sum_{m = all, n > 0} \right)
 \nonumber \\
 &\,& \times
 \langle n | \lambda_3 | m \rangle
 \langle m | (\gamma_5 + \Sigma_3)
 \left( \frac{\delta_n}{E_m - E_n} - \frac{1}{2} \,
 \delta_n^\prime \right)
 | n \rangle . \label{ga1e}
\end{eqnarray}
The $O(\Omega^1)$ contribution to the flavor non-singlet polarized 
distribution is a little more complicated. It generally consists of 
three terms, i.e. $O^{(1)}_{ \{A,B \} }, O^{(1)}_{ [A,B] }$ and 
$O^{(1)}_C$. Using the two identities, 
\begin{eqnarray}
 &\,& \sum_{m = all, M(n)} \frac{1}{E_m - E_n}
 [ \langle n | \lambda_b (\gamma_5 + \Sigma_3) \delta_n | m \rangle
 \langle m | \lambda_i | n \rangle \nonumber \\
 &\,& \hspace{25mm} + \,\langle n | \lambda_i | m \rangle
 \langle m | \lambda_b (\gamma_5 + \Sigma_3) \delta_n | n \rangle ] 
 \nonumber \\
 &=& \ \delta_{b 8} \delta_{i 3} \sum_{m = all, M(n)}
 \frac{1}{E_m - E_n} \,2 \,\langle n | \lambda_3 | m \rangle
 \langle m | \lambda_8 \,
 (\gamma_5 + \Sigma_3) \delta_n | n \rangle \nonumber \\
 &=& \ \frac{2}{\sqrt{3}} \,\delta_{b 8} \delta_{i 3} 
 \sum_{m = all, M(n)} \frac{1}{E_m - E_n} 
 \langle n | \lambda_3 | m \rangle
 \langle m | (\gamma_5 + \Sigma_3) \delta_n | n \rangle ,
\end{eqnarray}
and
\begin{eqnarray}
 &\,& \sum_{m = all, M(n)} \frac{1}{E_m - E_n}
 [ \langle n | \lambda_b (\gamma_5 + \Sigma_3 ) \delta_n | m \rangle
 \langle m | \lambda_K | n \rangle \nonumber \\
 &\,& \hspace{25mm} + \,\langle n | \lambda_K | m \rangle
 \langle m | \lambda_b (\gamma_5 + \Sigma_3) \delta_n | n \rangle ]
 \nonumber \\
 &=& 4 \,d_{3 K b} \sum_{m = all, M(n)} \frac{1}{E_m - E_n}
 \langle n | \lambda_4 | m \rangle 
 \langle m | \lambda_4 (\gamma_5 + \Sigma_3 ) 
 \delta_n | n \rangle ,
\end{eqnarray}
we obtain
\begin{eqnarray}
 O^{(1)}_{ \{A,B \} } &=& -\frac{2}{\sqrt{3}} \cdot \frac{1}{2}
 \{ D_{a 8}, J_3 \} \cdot M_N \frac{N_c}{4 I_1}
 \left( \sum_{m = all, n \leq 0} - \sum_{m = all, n > 0} \right) 
 \nonumber \\
 &\,& \hspace{20mm} \times \frac{1}{E_m - E_n} \,
 \langle n | \lambda_3 | m \rangle
 \langle m | (\gamma_5 + \Sigma_3 ) \delta_n | n \rangle \nonumber \\
 &\,& - 4 d_{3 K K} \frac{1}{2}
 \{ D_{a K}, J_K \} \cdot M_N \frac{N_c}{4 I_2}
 \left( \sum_{m = all, n \leq 0} - \sum_{m = all, n > 0} \right) 
 \nonumber \\
 &\,& \hspace{20mm} \times \frac{1}{E_m  - E_n} \,
 \langle n | \lambda_4 | m \rangle
 \langle m | \lambda_4 (\gamma_5 + \Sigma_3)
 \delta_n  | n \rangle .
\end{eqnarray}
Next, the $O^{(1)}_C$ term is given by
\begin{eqnarray}
 O^{(1)}_C &=& \ \frac{1}{2} \{ D_{a c}, J_i \}
 \frac{N_c}{8 I_1} \cdot \frac{d}{d x} 
 \left( \sum_{n \leq 0} - \sum_{n > 0} \right)
 \langle n | \{\lambda_c, \lambda_i \} (\gamma_5 + \Sigma_3) 
 \delta_n | n \rangle \nonumber \\ 
 &+& \ \frac{1}{2} \{ D_{a c}, J_K \} \frac{N_c}{8 I_2} 
 \cdot \frac{d}{d x} \left( \sum_{n \leq 0} - \sum_{n > 0} \right)
 \langle n | \{ \lambda_c, \lambda_K \} (\gamma_5 + \Sigma_3)
 \delta_n | n \rangle .
\end{eqnarray}
where $i$ runs from $1$ to $3$, while $K$ runs from $4$ to $7$.
To rewrite this term, we use two identities
\begin{eqnarray}
 &\,& \sum_{M(n)} \langle n | \{ \lambda_c, \lambda_i \} 
 (\gamma_5 + \Sigma_3 )
 \delta_n | n ) \nonumber \\
 &=& \frac{2}{\sqrt{3}} \,\delta_{c 8} \,\delta_{i 3} 
 \sum_{m = all, M(n)} 
 \langle n | \lambda_3 | m \rangle 
 \langle m | (\gamma_5 + \Sigma_3) \delta_n | n \rangle , \label{identa1}
\end{eqnarray}
and
\begin{eqnarray}
 &\,& \sum_{M(n)} \langle n | \{ \lambda_c. \lambda_K \} 
 (\gamma_5 + \Sigma_3) \delta_n | n \rangle \nonumber \\
 &=& 4 \,d_{3 c K} \sum_{m = all, M(n)}
 \langle n | \lambda_4 | m \rangle 
 \langle m | \lambda_4 (\gamma_5 + \Sigma_3)
 \delta_n | n \rangle , \label{identa2}
\end{eqnarray}
which will be proved in Appendix A.
We are then led to
\begin{eqnarray}
 O^{(1)}_C &=& \! \frac{2}{\sqrt{3}} \cdot \frac{1}{2}
 \{ D_{a 8}, J_3 \} \cdot \frac{N_c}{8 I_1} \frac{d}{d x}
 \left( \sum_{m = all,n \leq 0} - \sum_{m = all, n > 0} \right)
 \langle n | \lambda_3 | m \rangle \langle m | (\gamma_5 + \Sigma_3)
 \delta_n | n \rangle \nonumber \\
 &+& \!\!\! 4 \,d_{3 K K} \frac{1}{2} \{ D_{a K}^{(8)}, J_K \} \cdot
 \frac{N_c}{8 I_2} \frac{d}{d x} 
 \left( \sum_{m = all,n \leq 0} - \sum_{m = all, n > 0} \right)
 \langle n | \lambda_4 | m \rangle
 \langle m | \lambda_4 (\gamma_5 + \Sigma_3)
 \delta_n | n \rangle . \ \ \ \ \ 
\end{eqnarray}
Combining the $O^{(1)}_{ \{A,B \} }$ and $O^{(1)}_C$ terms,
we obtain
\begin{eqnarray}
 O^{(1)}_{ \{A,B\} } + O^{(1)}_C &=& \frac{2}{\sqrt{3}} \cdot
 \frac{1}{2} \{ D_{a 8}, J_3 \} \cdot e (x) \nonumber \\
 &+& 4 \,d_{3 K K} \cdot \frac{1}{2} \{ D_{a K}, J_K \} \cdot s (x) ,
\end{eqnarray}
where $e(x)$ is defined in (\ref{ga1e}), while $s(x)$ is defined by
\begin{eqnarray}
 s(x) &=& -M_N \frac{N_c}{4 I_2} 
 \left( \sum_{m = all, n \leq 0} - \sum_{m = all, n > 0} \right)
 \nonumber \\
 &\,& \hspace{10mm} \times \,
 \langle n | \lambda_4 | m \rangle 
 \langle m | \lambda_4 (\gamma_5 + \Sigma_3)
 \left(\frac{\delta_n}{E_m - E_n} - \frac{1}{2} \,
 \delta_n^\prime \right) 
 | n \rangle .
 \label{ga1s}
\end{eqnarray}
The remaining antisymmetric term, which is already familiar in 
the SU(2) CQSM, is given by                    
\begin{equation}
 O^{(1)}_{[A,B]} = -D_{a 3} \cdot h (x) ,
\end{equation}
with
\begin{eqnarray}
 h(x) &=& - \,i \,\varepsilon_{3 i j} \,M_N \frac{N_c}{8 I_1}
 \left( \sum_{m > 0, n \leq 0} + \sum_{m \leq 0, n > 0} \right)
 \frac{1}{E_m - E_n } \nonumber \\
 &\,& \times \,[ 
 \langle n | \lambda_j (\gamma_5 + \Sigma_3) \delta_n | m \rangle
 \langle m | \lambda_i | n \rangle
 - \langle n | \lambda_i | m \rangle
 \langle m | \lambda_j (\gamma_5 + \Sigma_3 )
 \delta_n | n \rangle ] .
\end{eqnarray}
The $O(\Omega^1)$ contribution to the flavor-nonsinglet 
polarized distribution then becomes
\begin{eqnarray}
 \Delta q^{(a)} (x : \Omega^1) &=& 
 \langle -D_{a 3} \rangle_{p \uparrow} \cdot \,h (x) 
 \nonumber \\
 &+& \langle 4 \sum_{K = 4}^7 d_{3 K K} \frac{1}{2}
 \{ D_{a K},J_K \} \rangle_{p \uparrow} \cdot \,s (x) \nonumber \\
 &+& \frac{2}{\sqrt{3}} \,
 \langle \frac{1}{2} \{ D_{a 8}, J_3 \} \rangle_{p \uparrow}
 \cdot \,e(x) \, .
\end{eqnarray}
At this stage, it would be convenient to summarize the complete 
forms of the unpolarized and longitudinally polarized distribution 
functions up to the first order in $\Omega$. First, for the unpolarized
distribution, the flavor singlet distribution is given by
\begin{equation} 
 q^{(0)} (x) \ = \ {\langle 1 \rangle}_p \cdot \,f (x) , \label{unpsing}
\end{equation}
whereas the flavor nonsinglet distributions $(a = 3 \ \mbox{or} \ 8)$
are given as
\begin{eqnarray}
 q^{(a)} (x) &=& 
 \langle \frac{D_{a 8}}{\sqrt{3}} \rangle_p \cdot \,f (x) 
 \nonumber \\
 &+& 
 \langle \sum_{i = 1}^3 \{ D_{a i}, R_i \} \rangle_p 
 \cdot \,k_1 (x) 
 \nonumber \\
 &+& 
 \langle \sum_{K = 4}^7 \{ D_{a K}, R_K \} \rangle_p
 \cdot \,k_2 (x) . \label{unpnsing}
\end{eqnarray}
Using the proton matrix elements of the relevant collective
operators :
\begin{eqnarray}
 \langle D_{3 8} / \sqrt{3} \rangle_p &=& \frac{1}{30}, \hspace{26mm} 
 \langle D_{88} / \sqrt{3} \rangle_p = 
 \frac{\sqrt{3}}{10} , \\
 \langle \sum_{i = 1}^3 \{ D_{3 i}, R_i \} \rangle_p 
 &=& \frac{7}{10}, \hspace{18mm}
 \langle \sum_{i = 1}^3 \{ D_{8 i}, R_i \} \rangle_p 
 = \frac{\sqrt{3}}{10} , \\
 \langle \sum_{K = 4}^7 \{ D_{3 K}, R_K \} \rangle_p 
 &=& \frac{1}{5} , \hspace{15mm}
 \langle \sum_{K = 4}^7 \{ D_{8 K}, R_K \} \rangle_p
 = \frac{3 \sqrt{3}}{5} ,
\end{eqnarray}
we finally arrive at
\begin{eqnarray}
 q^{(0)} (x) &=& f (x) , \\ \label{q0}
 q^{(3)} (x) &=& \frac{1}{30} \,f (x) + \frac{7}{10} \,k_1 (x)
 + \frac{1}{5} \,k_2 (x) , \\
 \frac{1}{\sqrt{3}} \,q^{(8)} (x) &=& \frac{1}{10} \,f (x) 
 + \frac{1}{10} \,k_1 (x) + \frac{3}{5} \,k_2 (x) .
\end{eqnarray}
These three distribution functions are enough to give the flavor 
decomposition of the unpolarized distribution functions :    
\begin{eqnarray}
 u (x) &=& \frac{1}{3} \,q^{(0)} (x) + \frac{1}{2} \,q^{(3)} (x)
 + \frac{1}{2 \sqrt{3}} \,q^{(8)} (x) , \\
 d (x) &=& \frac{1}{3} \,q^{(0)} (x) - \frac{1}{2} \,q^{(3)} (x) 
 + \frac{1}{2 \sqrt{3}} \,q^{(8)} (x) , \\
 s (x) &=& \frac{1}{3} \,q^{(0)} (x) \hspace{21mm}
 - \frac{1}{\sqrt{3}} \,q^{(8)} (x). \label{s}
\end{eqnarray}

The 1st-moment sum rules for the unpolarized distribution functions
are connected with the quark-number conservation laws.
The verification of them is therefore an important check of the
internal consistency of a theoretical formalism.
We first point out that the three basic distribution functions
of the model, i.e. $f(x), k_1 (x), k_2 (x)$, satisfy the sum rules
\begin{eqnarray}
 \int_{-1}^1 f(x) \,d x &=& 3, \\ \label{f1sum}
 \int_{-1}^1 k_1 (x) \,d x &=& 1, \\ \label{k1sum}
 \int_{-1}^1 k_2 (x) \,d x &=& 1. \label{k2sum}
\end{eqnarray}
Using (\ref{q0}) $\sim$ (\ref{s}) together with these sum rules,
it is an easy task to show that

\begin{eqnarray}
 \int_{-1}^1 q^{(0)} (x) \,d x &=& 3, \\
 \int_{-1}^1 q^{(3)} (x) \,d x &=& 1, \\
 \int_{-1}^1 q^{(8)} (x) \,d x &=& 1,
\end{eqnarray}
and
\begin{eqnarray}
 \int_{-1}^1 u (x) \,d x &=& \int_0^1 \,[ u (x) - \bar{u} (x) ] \,d x
 \ = \ 2, \\
 \int_{-1}^1 d (x) \,d x &=& \int_0^1 \,[ d (x) - \bar{d} (x) ] \,d x
 \ = \ 1, \\
 \int_{-1}^1 s (x) \,d x &=& \int_0^1 \,[ s (x) - \bar{s} (x) ] \,d x
 \ = \ 0,
\end{eqnarray}
which are just the desired quark-number conservation laws.

Incidentally, the unpolarized distribution functions in the SU(2)
CQSM are given in the following form :
\begin{eqnarray}
 u (x) &=& \frac{1}{2} \,q^{(0)} (x) + \frac{1}{2} \,q^{(3)} (x) , \\
 d (x) &=& \frac{1}{2} \,q^{(0)} (x) - \frac{1}{2} \,q^{(3)} (x) , \\
 s (x) &=& \hspace{6mm} 0 ,
\end{eqnarray}
where
\begin{eqnarray} 
 q^{(0)} (x) &=& f (x) , \\
 q^{(3)} (x) &=& k_1 (x) .
\end{eqnarray}
with $f(x)$ and $k_1 (x)$ being the same functions as appear in the
SU(3) CQSM.

Next, the $O(\Omega^0 + \Omega^1)$ contributions to the longitudinally
polarized distribution functions can be summarized as 
\begin{equation}
 \Delta q^{(0)} (x) \ = \ 
 {\langle 2 J_3 \rangle}_{p \uparrow} \cdot \,e (x) , \label{lgpsing}
\end{equation}
for the flavor-singlet distributions, and
\begin{eqnarray}
 \Delta q^{(a)} (x) &=& 
 {\langle -D_{a 3} \rangle}_{p \uparrow} \cdot \,
 (g (x) + h (x) ) \nonumber \\
 &+& \langle 4 \sum_{K = 4}^7 d_{3 K K} \frac{1}{2}
 \{ D_{a K}, J_K \} \rangle_{p \uparrow} \cdot \,s (x) \nonumber \\
 &+& \frac{2}{\sqrt{3}} \,
 {\langle \frac{1}{2} \{ D_{a 8}, J_3 \} \rangle} \cdot \,e (x) ,
 \label{lgpnsing}
\end{eqnarray}
for the nonsinglet distributions.
Using the matrix elements of the relevant collective space operators
between the spin-up proton state,      
\begin{eqnarray}
 \langle -D_{33} \rangle_{p \uparrow} &=& \frac{7}{30}, \hspace{15mm}
 \langle -D_{83} \rangle_{p \uparrow} = 
 \frac{\sqrt{3}}{30} ,\nonumber \\
 \langle 4 \sum_{K = 4}^7 d_{3 K K} D_{3 K} J_K \rangle_{p \uparrow}
 &=& \frac{7}{15} , \hspace{15mm}
 \langle 4 \sum_{K = 4}^7 d_{3 K K} D_{8 K} J_K \rangle_{p \uparrow}
 = \frac{\sqrt{3}}{15} , \nonumber \\
 \langle D_{38} J_3 \rangle_{p \uparrow}
 &=& \frac{\sqrt{3}}{60} , \hspace{15mm}
 \langle D_{88} J_3 \rangle_{p \uparrow}
 = \frac{\sqrt{3}}{20} ,
\end{eqnarray}
we obtain
\begin{eqnarray}
 \Delta q^{(0)} (x) &=& e (x) ,\\
 \Delta q^{(3)} (x) &=& \frac{1}{30} e (x) 
 + \frac{7}{30} \,(g (x) + h (x) )
 + \frac{7}{15} s (x) , \\
 \frac{1}{\sqrt{3}} \,\Delta q^{(8)} (x) 
 &=& \frac{1}{10} e (x) 
 + \frac{1}{30} \,(g (x) + h (x) ) 
 + \frac{1}{15} s (x)  .
\end{eqnarray}
In terms of these 3 functions, the longitudinally polarized 
distribution functions with each flavor are given by   
\begin{eqnarray}
 \Delta u (x) &=& \frac{1}{3} \,\Delta q^{(0)} (x) 
 + \frac{1}{2} \,\Delta q^{(3)} (x) 
 + \frac{1}{2 \sqrt{3}} \,\Delta q^{(8)} (x) , \\
 \Delta d (x) &=& \frac{1}{3} \,\Delta q^{(0)} (x)
 - \frac{1}{2} \,\Delta q^{(3)} (x) 
 + \frac{1}{2 \sqrt{3}} \,\Delta q^{(8)} (x) , \\
 \Delta s (x) &=& \frac{1}{3} \,\Delta q^{(0)} (x) 
 \hspace{23mm} - \,\frac{1}{\sqrt{3}} \,\,\Delta q^{(8)} (x) .
\end{eqnarray}
For comparison, we also show the corresponding theoretical formulas
obtained within the framework of the SU(2) CQSM : 
\begin{eqnarray}
 \Delta u (x) &=& \frac{1}{2} \,\Delta q^{(0)} (x) 
 + \frac{1}{2} \,\Delta q^{(3)} (x) ,\\
 \Delta d (x) &=& \frac{1}{2} \,\Delta q^{(0)} (x) 
 + \frac{1}{2} \,\Delta q^{(3)} (x) , \\
 \Delta s (x) &=& \hspace{6mm} 0 ,
\end{eqnarray}
where
\begin{eqnarray}
 \Delta q^{(0)} (x) &=& e (x) , \\
 \Delta q^{(3)} (x) &=& \frac{1}{3} \,(g (x) + h(x) ) .
\end{eqnarray}
We recall here that, as a consequence of the new operator ordering 
procedure adopted in the present paper, one noteworthy difference
with the previous treatment arises, concerning the $O(\Omega^1)$
contribution to the isovector distribution $\Delta q^{(3)} (x)$.
Namely, the $[ \Delta u (x) - \Delta d (x) ]^{(1)}_{B^\prime + C}$
term in eq.(114) of Ref.\cite{WK99} is totally absent
in the new formulation here. 
We shall numerically check that the effect of this change on the 
final predictions for the longitudinally polarized distributions 
is very small.

Similarly as in the case of the unpolarized distributions, we can
write down the 1st-moment sum rules also longitudinally polarized
distributions. No exact conservation law follows from these sum
rules, however. As a matter of course, this does not mean there
is no useful sum rule for the spin-dependent distributions.
For example, the celebrated Bjorken sum rule \cite{BJ66},\cite{BJ70}
for the isovector part of the longitudinally polarized
distribution functions has an
important phenomenological significance, although it is not a sort
of relation which gives an exact conservation laws for some quantum
numbers.

\subsection{$\Delta m_s$ corrections to PDF}

Our strategy for estimating the SU(3) symmetry breaking effects
is to use the first order perturbation theory in $\Delta m_s$, i.e.
the mass difference between the $s$- and $u,d$-quarks. There
are several such corrections that are all first order in 
$\Delta m_s$. The first comes from (\ref{delms}) containing the
SU(3) symmetry breaking part of the effective Dirac hamiltonian 
$\Delta H_{\alpha \beta}$. Following Ref.~\cite{BPG96}, this SU(3) 
symmetry breaking correction is hereafter referred to as the 
``dynamical $\Delta m_s$ correction''.
The second correction originates from the term (\ref{omega1}),
which is first order in $\Omega$, if it is combined with the
quantization rule (\ref{qrulems}) of the SU(3) collective rotation.
In fact, one can easily convince that the replacement.
\begin{eqnarray}
 \Omega_{\alpha \beta} \ = \ \frac{1}{2} \Omega_a 
 (\lambda_a)_{\alpha \beta} 
 &\rightarrow& \ \frac{1}{2} \left( \frac{J_i}{I_1} 
 + \frac{2}{\sqrt{3}} \Delta m_s 
 \frac{K_1}{I_1} D_{8i} \right) (\lambda_i)_{\alpha \beta} \nonumber \\
 &+& \frac{1}{2} \left( \frac{J_K}{I_2} 
 + \frac{2}{\sqrt{3}} \Delta m_s
 \frac{K_2}{I_2} D_{8 K} \right) (\lambda_K)_{\alpha \beta} ,
\end{eqnarray}
brings about terms proportional to the mass difference 
$\Delta m_s$. This SU(3) symmetry breaking correction, which 
comes from the $\Delta m_s$ correction to the SU(3) quantization 
rule, will be called the ``kinematical $\Delta m_s$ correction". 
The third correction is brought about by the mixing of the SU(3)
irreducible representations, describing the baryon states as 
collective rotational states. Since this mixing occurs also at 
the first order in $\Delta m_s$, we must take account of it. 
This last SU(3) symmetry breaking correction will be called the
``representation-mixing $\Delta m_s$ correction". In the following,
we shall treat these three corrections in order.
The answer will be given in the form :
\begin{equation}
 q(x ; \Delta m_s) = \int \Psi_{Y T T_3 ; J J_3}^{(n)*} [\xi_A] \,
 O^{(\Delta m_s)} [\xi_A] \,\Psi_{Y T T_3 ; J J_3} [\xi_A] d \xi_A ,
\end{equation}
where the effective collective space operator consists of three parts :
\begin{equation}
 O^{(\Delta m_s)} [\xi_A] = O^{(\Delta m_s)}_{dyn} + 
 O^{(\Delta m_s)}_{kin} + O^{(\Delta m_s)}_{rep} .
\end{equation}
First to evaluate $O_{dyn}^{(\Delta m_s)}$ by using (\ref{delms}),
the ordering
\begin{eqnarray}
   &\,& \!\!\!\!\! \Delta H_{\alpha \beta} (z'_0) \,
   {(A^\dagger (0) O_a A(z_0))}_{\gamma \delta} \nonumber \\
   &\longrightarrow& [ \theta (z'_0, 0, z_0) + \theta (z'_0, z_0, 0) ]
   \,\Delta H_{\alpha \beta} \,\tilde{O}_{\gamma \delta} \nonumber \\
   &+& [ \theta (0, z_0, z'_0) + \theta (z_0, 0, z'_0) ] \,
   \tilde{O}_{\gamma \delta} \Delta H_{\alpha \beta}  \nonumber \\
   &+& \ \theta(0, z'_0, z_0) \,\frac{1}{2}
   \{ \Delta H_{\alpha \beta}, \tilde{O}_{\gamma \delta} \} 
   \nonumber \\
   &+& \ \theta(z_0, z'_0, 0) \,\frac{1}{2}
   \{ \Delta H_{\alpha \beta}, \tilde{O}_{\gamma \delta} \} ,
\end{eqnarray}
is used in conformity with the rule (\ref{ordering}).
After carrying out the integration over the variables
$\mbox{\boldmath $R$}, \mbox{\boldmath $z$}^\prime, z_0^\prime$
and over $z_0$, we are led to the following answer for the
dynamical $\Delta m_s$ correction :
\begin{eqnarray}
 O^{(\Delta m_s)} &=& - M_N \frac{N_c}{2} 
 \left( \sum_{m = all, n \leq 0} - \sum_{m = all, n > 0} \right)
 \frac{1}{E_m - E_n} \nonumber \\
 &\,& \times \,\left[ \langle n \vert \Delta H \vert m \rangle
 \langle m \vert \tilde{O} \delta_n \vert n \rangle + 
 \langle n \vert \tilde{O} \delta_n \vert m \rangle
 \langle m \vert \Delta H \vert n \rangle \right] ,
\end{eqnarray}
where $\tilde{O} = A^\dagger \lambda_a A \bar{O}$ and
\begin{eqnarray}
 \Delta H &=& \Delta m_s \gamma^0 A^\dagger \left(
 \frac{1}{3} - \frac{1}{\sqrt{3}} \lambda_8 \right) A
 \ = \ \Delta m_s \gamma^0 \left(
 \frac{1}{3} - \frac{1}{\sqrt{3}} D_{8c} \lambda_c \right) .
\end{eqnarray}
In deriving the above equation, we have used the fact that the
collective operators contained in $\Delta H$ and
$\tilde{O}$ commute each other.

In the case of flavor singlet unpolarized distribution, the above
general formula gives
\begin{eqnarray}
 O^{(\Delta m_s)}_{dyn} &=& -M_N \frac{N_c}{2} 
 \left(\sum_{m = all, n \leq 0} - \sum_{m = all, n > 0} \right)
 \frac{1}{E_m - E_n} \cdot \frac{1}{3} \Delta m_s \nonumber \\
 &\,& \hspace{10mm} \times \,
 [ \langle n | (1 - \sqrt{3} D_{8 c} \lambda_c)
 \gamma^0 | m \rangle 
 \langle m | (1 + \gamma^0 \gamma^3) \delta_n | n \rangle \nonumber \\
 &\,& \hspace{10mm} 
 + \,\langle n | (1 + \gamma^0 \gamma^3) \delta_n | m \rangle
 \langle m | (1 - \sqrt{3} D_{8 c} \lambda_c) \gamma^0 | n \rangle ]
 \nonumber \\
 &=& -\frac{1}{3} \Delta m_s \cdot M_N \frac{N_c}{2} 
 \left( \sum_{m = all, n \leq 0} - \sum_{m = all, n > 0} \right)
 \frac{1}{E_m - E_n} \nonumber \\
 &\,& \hspace{10mm}
 \times \,[ \langle n | (1 - \sqrt{3} D_{88} \lambda_8 ) \gamma^0 
 | \rangle \langle m | (1 + \gamma^0 \gamma^3 ) 
 \delta_n | n \rangle \nonumber \\
 &\,& \hspace{10mm} 
 + \,\langle n | (1 + \gamma^0 \gamma^3 ) \delta_n | m \rangle 
 \langle m | (1 - \sqrt{3} D_{88} \lambda_8 ) 
 \gamma^0 | n \rangle ] . \label{dynms}
\end{eqnarray}
Using the generalized hedgehog symmetry, we therefore arrive at
\begin{equation}
 q^{(0)} (x : \Delta m_s^{dyn} ) = -\frac{4}{3} \,
 \langle 1 - D_{88} \rangle \cdot \Delta m_s I_1 \cdot 
 \tilde{k_0} (x) ,
\end{equation}
with 
\begin{eqnarray}
 \tilde{k_0} (x) &=& \frac{1}{I_1} \cdot \frac{N_c}{4} \,
 \left(\sum_{m = all, n \leq 0} - \sum_{m = all, n > 0} \right) 
 \frac{1}{E_m - E_n} \,
 \langle n | \gamma^0 | m \rangle
 \langle m | (1 + \gamma^0 \gamma^3 ) \delta_n | n \rangle .
 \label{kw0}
\end{eqnarray}
Next we turn to the flavor nonsinglet unpolarized distributions 
$(a = 3 \ \mbox{or} \ 8)$. The general formula (\ref{dynms}) gives
\begin{eqnarray}
 O^{(\Delta m_s)}_{dyn} &=& -M_N \,\frac{N_c}{2} \,
 \left(\sum_{m = all, n \leq 0} - \sum_{m = all, n > 0} \right)
 \frac{1}{E_m - E_n} \cdot \frac{1}{3} \Delta m_s \nonumber \\
 &\,& \hspace{10mm} 
 \times \,[ \langle n | (1 - \sqrt{3} D_{8 c} \lambda_c )
 \gamma^0 | m \rangle \langle m | D_{a b} \lambda_b 
 ( 1 + \gamma^0 \gamma^3) \delta_n | n \rangle \\
 &\,& \hspace{10mm}
 + \,\langle n | D_{a b} \lambda_b (1 + \gamma^0 \gamma^3 ) 
 \delta_n | m \rangle \langle m | 
 ( 1 - \sqrt{3} D_{8 c} \lambda_c ) \gamma^0 | n \rangle  ] 
 \nonumber \\
 &=& - \,\frac{1}{3} \,\Delta m_s \cdot M_N \frac{N_c}{2} \,
 \left(\sum_{m = all, n \leq 0} - \sum_{m = all, n > 0} \right)
 \frac{1}{E_m - E_n} \nonumber \\
 &\,& \times \,\{ D_{a b} [ \langle n | \gamma^0 | m \rangle
 \langle m | \lambda_b (1 + \gamma^0 \gamma^3 ) 
 \delta _n | n \rangle + 
 \langle n | \lambda_b (1 + \gamma^0 \gamma^3 ) \delta_n | m 
 \rangle \langle  m | \gamma^0 | n \rangle ] \nonumber \\
 &\,& \hspace{10mm}
 - \,\sqrt{3} D_{8 c} D_{a b} \langle n | \lambda_c \gamma^0 
 | m \rangle \langle m | \lambda_b (1 + \gamma^0 \gamma^3 ) 
 \delta_n | n \rangle \nonumber \\
 &\,& \hspace{10mm}
 - \,\sqrt{3} D_{a b} D_{8 c} \langle n | \lambda_b 
 (1 + \gamma^0 \gamma^3 )\delta _n | m \rangle  \langle m 
 | \lambda_c \gamma^0 | n \rangle \} . \label{dynmsns}
\end{eqnarray}
It is easy to show that the contribution of the first term of
(\ref{dynmsns}), (i.e. the term proportional to $D_{a b}$), to
$q^{(a)} ( x, \Delta m_s^{dyn} )$ is given by 
\begin{equation}
 -\frac{4 \Delta m_s I_1}{3} \,\langle \frac{D_{a 8}}{\sqrt{3}} 
 \rangle_{p \uparrow} \cdot \tilde{k_0} (x)
\end{equation}
with $\tilde{k_0} (x)$ given by (\ref{kw0}). The manipulation of the 
remaining two terms is a little more complicated. First, we notice 
that, since $D_{8 c}$ and $D_{a b}$ commute, we can write as
\begin{eqnarray}
 &\,& D_{8 c} D_{a b} \,\langle n | \lambda_c \gamma^0 | m \rangle
 \langle m | \lambda_b (1 + \gamma^0 \gamma^3 ) \delta_n | n \rangle
 \nonumber\\
 &+& D_{a b} D_{8 c} \,\langle n | \lambda_b (1 + \gamma^0 \gamma^3 )
 \delta_n | m \rangle \langle m | \lambda_c \gamma^0 | n \rangle
 \nonumber \\
 &=& \frac{1}{2} \{ D_{a b}, D_{8 c} \} \,[ \langle n | \lambda_b 
 (1 + \gamma^0 \gamma^3 ) \delta_n | m \rangle
 \langle m | \lambda_c \gamma^0 | n \rangle \nonumber \\
 &\,& \hspace{20mm} + \langle n | \lambda_c \gamma^0 | m \rangle
 \langle m | \lambda_b (1 + \gamma^0 \gamma^3 ) 
 \delta_n | n \rangle ] \, .
\end{eqnarray}
We now consider the two parts separately. For the parts where 
the indices $b$ and $c$ run from $1$ to $3$, the diagonal matrix 
element of $D_{a b} D_{8 c}$ between the spin-up proton state can 
be expressed 
\begin{eqnarray}
 \langle D_{a b} D_{8 c} \rangle^{b,c = 1,2,3 \ part}_{p \uparrow}
 &=& \langle D_{a 3} D_{8 c} \rangle_{p \uparrow} \,\,
 \delta_{b,3} \,\delta_{c,3} \nonumber \\
 &+& \frac{1}{4} \langle D_{a, 1 + i 2} D_{8,1 - i 2} \,
 \rangle_{p \uparrow} \,\,\delta_{b, 1 - i 2} \,\delta_{c, 1 + i 2}
 \nonumber \\
 &+& \frac{1}{4} \langle D_{a, 1 - i 2} D_{8, 1 + i 2} \,
 \rangle_{p \uparrow} \,\,\delta_{b, 1 + i 2} \,\delta_{c, 1 - i 2} .
\end{eqnarray}
Noting the equalities
\begin{equation}
 \langle D_{a, 1 + i 2} D_{8, 1 - i 2} \rangle_{p \uparrow}
 = \langle D_{a, 1 - i 2} D_{8, 1 + i 2} \rangle_{p \uparrow}
 = 2 \,\langle D_{a 3} D_{8 3} \rangle _{p \uparrow} ,
\end{equation}
we can prove that 
\begin{equation}
 \langle D_{a b} D_{8 c} \rangle^{b,c = 1,2,3\ part}_{p \uparrow}
 = \langle D_{a 3} D_{8 3} \rangle _{p \uparrow} \,
 (\delta_{b,1} \delta_{c,1} + \delta_{b,2} \delta_{c,2} 
 + \delta_{b,3} \delta_{c,3} ) .
\end{equation}
Using the similar relation for the product of operators 
$D_{8 c} D_{a b}$, we then get
\begin{eqnarray}
 &\,& {\langle \{ D_{a b} D_{8 c} \} 
 \rangle}_{p \uparrow}^{b,c = 1,2,3 part}
 \ = \ \frac{1}{3} \,\langle \sum_{i = 1}^3 \{ D_{a i} D_{8 i} \}
 \rangle_{p \uparrow} \,\,(\delta_{b,1} \delta_{c,1} 
 + \delta_{b,2} \delta_{c,2} + \delta_{b,3} \delta_{c,3} ) .
\end{eqnarray}
This relation is then used to derive the equality :
\begin{eqnarray}
 &\,& \sum_{b,c = 1}^3 \frac{1}{2} {\langle \{ D_{a b} D_{8 c} \} 
 \rangle}_{p \uparrow} \frac{N_c}{2}
 \left( \sum_{m = all, n \leq 0} - \sum_{m = all, n > 0} \right)
 \frac{1}{E_m - E_n} \nonumber \\
 &\,& \times \,[ \langle n | \lambda_b (1 + \gamma^0 \gamma^3 )
 \delta_n | m \rangle \langle m | \lambda_c \gamma^0 | n \rangle
 + \,\langle n | \lambda_c \gamma^0 | m \rangle
 \langle m | \lambda_b (1 + \gamma^0 \gamma^3 ) 
 \delta_n | n \rangle ] \nonumber \\
 &=& \langle \sum_{i = 1}^3 \{ D_{a i}, D_{8 i} \} \rangle_{p \uparrow}
 \cdot \frac{1}{3} \sum_{j = 1}^3 \frac{N_c}{2} 
 \langle(\sum_{m = all, n \leq 0} - \sum_{m = all, n > 0} \rangle)
 \nonumber \\
 &\,& \hspace{35mm} \times \frac{1}{E_m - E_n} \,
 \langle n | \lambda_j \gamma^0 | m \rangle
 \langle m | \lambda_j (1 + \gamma^0 \gamma^3 ) \delta_n | n \rangle .
\end{eqnarray}
Next, for the parts where $b$ and $c$ run from $4$ to $7$, 
we use the identities
\begin{eqnarray}
 &\,& \sum_{m = all, M(n)} \langle n | 
 \lambda_4 \gamma^0 | m \rangle
 \langle m | \lambda_4 (1 + \gamma^0 \gamma^3 ) \delta_n | n \rangle
 \nonumber \\
 &=& \sum_{m = all, M(n)} \langle n | \lambda_5 \gamma^0 | m \rangle
 \langle m | \lambda_5 (1 + \gamma^0 \gamma^3 ) \delta_n | n \rangle
 \nonumber \\
 &=& \sum_{m = all, M(n)} \langle n | \lambda_6 \gamma^0 | m \rangle
 \langle m | \lambda_6 (1 + \gamma^0 \gamma^3 ) \delta_n | n \rangle
 \nonumber \\
 &=& \sum_{m = all, M(n)} \langle n | \lambda_7 \gamma^0 | m \rangle
 \langle m | \lambda_7 (1 + \gamma^0 \gamma^3 ) \delta_n | n \rangle
\end{eqnarray}
Using these relations, we find that
\begin{eqnarray}
 &\,& \!\!\! \sum_{b,c = 4}^7 \frac{1}{2} \,
 {\langle \{ D_{a b},D_{8 c} \} 
 \rangle}_{p \uparrow} \,\frac{N_c}{2} 
 \left( \sum_{m = all,n \leq 0} - \sum_{m = all, n > 0} \right)
 \frac{1}{E_m - E_n} \nonumber \\
 &\,& \ \ \times \,[ \langle n | \lambda_b (1 + \gamma^0 \gamma^3 ) 
 \delta_n | m \rangle \langle m | \lambda_c \gamma^0 | n \rangle
 + \langle n | \lambda_c \gamma^0 | m \rangle
 \langle m | \lambda_b (1 + \gamma^0 \gamma^3 ) 
 \delta_n | n \rangle ] \nonumber \\
 &=& \langle \sum_{K = 4}^7 \{ D_{a K},D_{8 K} \} 
 \rangle_{p \uparrow} \,
 \frac{N_c}{2} \left(\sum_{m = all, n \leq 0} 
 - \sum_{m = all, n > 0} \right) \nonumber \\ 
 &\,& \hspace{20mm} \times \,\frac{1}{E_m - E_n} \langle n | 
 \lambda_4 \gamma^0 | m \rangle
 \langle m | \lambda_4 (1 + \gamma^0 \gamma^3 ) 
 \delta_n | n \rangle ,
\end{eqnarray}
Now combining the above three contributions, the dynamical $\Delta m_s$
corrections to the flavor nonsinglet unpolarized distributions can be
written in the form :
\begin{eqnarray}
 q^{(a)} (x ; \Delta m_s^{dyn}) &=&
 - \frac{4 \Delta m_s I_1}{3} \cdot
 \langle \frac{D_{a8}}{\sqrt{3}} \rangle_{p \uparrow} 
 \cdot \,\tilde{k}_0 (x) \nonumber \\
 &+& \frac{2 \Delta m_s I_1}{\sqrt{3}} \cdot
 \langle \sum_{i=1}^3 \{ D_{ai}, D_{8i} \} \rangle_{p \uparrow}
 \cdot \,\tilde{k}_1 (x) \nonumber \\
 &+& \frac{2 \Delta m_s I_2}{\sqrt{3}} \cdot
 \langle \sum_{K=4}^7 \{ D_{aK}, D_{8K} \} \rangle_{p \uparrow}
 \cdot \,\tilde{k}_2 (x) ,
\end{eqnarray}
where $\tilde{k}_0 (x)$ is already defined in (\ref{kw0}), while
\begin{eqnarray}
 \tilde{k}_1 (x) &=& M_N \frac{N_c}{4 I_1} \cdot \frac{1}{3}
 \sum_{j=1}^3 
 \left( \sum_{m = all, n \leq 0} - \sum_{m = all, n >0} \right) 
 \nonumber \\
 &\,& \hspace{15mm} \times \,\frac{1}{E_m - E_n} 
 \langle n \vert \lambda_j \gamma^0 \vert m \rangle
 \langle m \vert \lambda_j (1 + \gamma^0 \gamma^3) 
 \delta_n \vert n \rangle, \\
 \tilde{k}_2 (x) &=& M_N \frac{N_c}{4 I_2} \cdot
 \left( \sum_{m = all, n \leq 0} - \sum_{m = all, n >0} \right) 
 \nonumber \\
 &\,& \hspace{15mm} \times \,\frac{1}{E_m - E_n} 
 \langle n \vert \lambda_4 \gamma^0 \vert m \rangle
 \langle m \vert \lambda_4 (1 + \gamma^0 \gamma^3) 
 \delta_n \vert n \rangle.
\end{eqnarray}
Next, we consider the longitudinally polarized distributions. The flavor
singlet part is easily obtained in the form : 
\begin{equation}
 \Delta q^{(0)} (x ; \Delta m_s^{dyn}) = - \frac{4 \Delta m_s I_1}{\sqrt{3}}
 \cdot {\langle D_{83} \rangle}_{p \uparrow} \cdot \tilde{e} (x) ,
\end{equation}
with
\begin{eqnarray}
 \tilde{e} (x) &=& - M_N \frac{N_c}{4 I_1}
 \left( \sum_{m = all, n \leq 0} - \sum_{m = all, n > 0} \right) \nonumber \\
 &\,& \hspace{10mm} \times \frac{1}{E_m - E_n}
 \langle n \vert \lambda_3 \gamma^0 \vert m \rangle
 \langle m \vert (\gamma_5 + \Sigma_3) \delta_n \vert n \rangle .
 \label{ga1ew}
\end{eqnarray}
The flavor nonsinglet part is again a slightly more complicated. From
the general formula (\ref{dynmsns}), we get
\begin{eqnarray}
 O_{dyn}^{(\Delta m_s)} &=& - M_N \frac{N_c}{2}
 \left( \sum_{m = all, n \leq 0} - \sum_{m = all, n > 0} \right)
 \frac{1}{E_m - E_n} \frac{1}{3} \Delta m_s \nonumber \\
 &\,& \hspace{15mm} \times \left[
 \langle n \vert (1 - \sqrt{3} D_{8c} \lambda_c) \gamma^0 \vert m \rangle
 \langle m \vert D_{ab} \lambda_b (\gamma_5 + \Sigma_3) \delta_n
 \vert n \rangle \right. \nonumber \\
 &\,& \hspace{15mm} \left. 
 + \,\langle n \vert D_{ab} \lambda_b (\gamma_5 + \Sigma_3) \delta_n
 \vert n \rangle
 \langle m \vert (1 - \sqrt{3} D_{8c} \lambda_c) \gamma^0 \vert n \rangle
 \right] \nonumber \\
 &=& - \frac{1}{3} \Delta m_s \cdot M_N \frac{N_c}{2}
 \left( \sum_{m = all, n \leq 0} - \sum_{m = all, n > 0} \right)
 \frac{1}{E_m - E_n} \nonumber \\
 &\,& \times \, \{
 D_{ab} \,[
 \langle n | \gamma^0 | m \rangle
 \langle m | \lambda_b (\gamma_5 + \Sigma_3) \delta_n | n \rangle +
 \langle n | \lambda_b (\gamma_5 + \Sigma_3) \delta_n | m \rangle
 \langle m | \gamma^0 | n \rangle ] \nonumber \\
 &\,& \hspace{15mm} - \sqrt{3} D_{8c} D_{ab} 
 \langle n \vert \lambda_c \gamma^0 \vert m \rangle
 \langle m \vert \lambda_b (\gamma_5 + \Sigma_3) \delta_n \vert n \rangle
 \nonumber \\
 &\,& \hspace{15mm} - \sqrt{3} D_{ab} D_{8c} 
 \langle n \vert \lambda_b (\gamma_5 + \Sigma_3) \delta_n \vert m \rangle
 \langle m \vert \lambda_c \gamma^0 \vert n \rangle \} .
\end{eqnarray}
Similarly as before, the contribution of the first term (proportional to
$D_{ab}$) to $\Delta q^{(0)} (x ; \Delta m_s^{dyn})$ is found to be
\begin{equation}
 \frac{4 \Delta m_s I_1}{3} \,{\langle D_{a3} \rangle}_{p \uparrow} \cdot
 \tilde{e} (x) ,
\end{equation}
with $\tilde{e} (x)$ given by eq.(\ref{ga1ew}).
On the other hand, the remaining
two terms can be rewritten in the form : 
\begin{eqnarray}
 &\,& \frac{1}{\sqrt{3}} \,\Delta m_s \cdot M_N \frac{N_c}{2}
 \left( \sum_{m = all, n \leq 0} - \sum_{m = all, n > 0} \right)
 \frac{1}{E_m - E_n} \nonumber \\
 &\,& \times \frac{1}{2} \{ D_{ab}, D_{8c} \}
 \left[ 
 \langle n \vert \lambda_b (\gamma_5 + \Sigma_3) \delta_n \vert m \rangle
 \langle m \vert \lambda_c \gamma^0 \vert n \rangle + 
 \langle n \vert \lambda_c \gamma^0 \vert m \rangle
 \langle m \vert \lambda_b (\gamma_5 + \Sigma_3) \delta_n \vert n \rangle
 \right] . \ \ \ \ \ 
\end{eqnarray}
First by confining to the terms in which either or both of $b$ and $c$
run from 1 to 3, there are only two possibilities to survive, i.e.
$b = 8, c = 3$ or $b = 3, c = 8$. The contributions of these terms to
$q^{(a)} (x ; \Delta m_s^{dyn})$ are found to be
\begin{eqnarray}
 &\,& \!\!\!\! \frac{4 \Delta m_s I_1}{3}
 {\langle D_{a3} D_{88} \rangle}_{p \uparrow} M_N \frac{N_c}{4 I_1}
 \left( \sum_{m = all, n \leq 0} - \sum_{m = all, n > 0} \right)
 \frac{1}{E_m - E_n}
 \langle n \vert \gamma^0 \vert m \rangle
 \langle m \vert \lambda_3 (\gamma_5 + \Sigma_3) 
 \delta_n \vert n \rangle
 \nonumber \\
 &+& \!\!\! \frac{4 \Delta m_s I_1}{3}
 {\langle D_{a8} D_{83} \rangle}_{p \uparrow} M_N \frac{N_c}{4 I_1}
 \left( \sum_{m = all, n \leq 0} - \sum_{m = all, n > 0} \right)
 \frac{1}{E_m - E_n}
 \langle n \vert \lambda_3 \gamma^0 \vert m \rangle
 \langle m \vert (\gamma_5 + \Sigma_3) \delta_n \vert n \rangle .
 \ \ \ \ \ \ \ 
\end{eqnarray}
In order to evaluate the remaining contributions in which $b$ and $c$
run from 4 to 7, we use the identity :
\begin{eqnarray}
 &\,& \sum_{m = all, M(n)} \langle n \vert \lambda_4 
 \gamma^0 \vert m \rangle
 \langle m \vert \lambda_4 (\gamma_5 + \Sigma_3) 
 \delta_n \vert n \rangle
 \nonumber \\
 &=& \sum_{m = all, M(n)} \langle n \vert \lambda_5 
 \gamma^0 \vert m \rangle
 \langle m \vert \lambda_5 (\gamma_5 + \Sigma_3) 
 \delta_n \vert n \rangle
 \nonumber \\
 &=& - \sum_{m = all, M(n)} \langle n \vert \lambda_6 
 \gamma^0 \vert m \rangle
 \langle m \vert \lambda_6 (\gamma_5 + \Sigma_3) 
 \delta_n \vert n \rangle
 \nonumber \\
 &=& - \sum_{m = all, M(n)} \langle n \vert \lambda_7 
 \gamma^0 \vert m \rangle
 \langle m \vert \lambda_7 (\gamma_5 + \Sigma_3) 
 \delta_n \vert n \rangle ,
\end{eqnarray}
together with the familiar relation : 
\begin{equation}
 d_{344} = d_{355} = - d_{366} = - d_{377} = \frac{1}{2} .
\end{equation}
This enables us to express the corresponding contribution to
$q^{(a)} (x ; \Delta m_s^{dyn})$ in the following form :
\begin{eqnarray}
 &\,& \frac{2 \Delta m_s I_2}{\sqrt{3}} \,
 \langle 4 \sum_{K=4}^7 d_{3KK} D_{aK} D_{8K} \rangle_{p \uparrow} \,
 M_N \frac{N_c}{4 I_2}
 \left( \sum_{m = all, n \leq 0} - \sum_{m = all, n > 0} \right) 
 \nonumber \\
 &\,& \hspace{40mm} \times \frac{1}{E_m - E_n} \,
 \langle n \vert \lambda_4 \gamma^0 \vert m \rangle
 \langle m \vert \lambda_4 (\gamma_5 + \Sigma_3) 
 \delta_n \vert n \rangle .
\end{eqnarray}
Now, by collecting the various terms explained above, the dynamical
$\Delta m_s$ correction to the flavor nonsinglet longitudinally polarized
distribution functions can be expressed as
\begin{eqnarray}
 \Delta q^{(a)} (x ; \Delta m_s^{dyn}) &=&
 \frac{4 \Delta m_s I_1}{3} \,\,
 {\langle D_{a3} (1 - D_{88}) \rangle}_{p \uparrow}
 \cdot \tilde{f} (x) \nonumber \\
 &-& \frac{4 \Delta m_s I_1}{3} \,\,
 {\langle D_{a8} D_{83} \rangle}_{p \uparrow}
 \cdot \tilde{e} (x) \nonumber \\
 &-& \frac{2 \Delta m_s I_2}{\sqrt{3}} \,\,
 \langle 4 \sum_{K=4}^7 d_{3KK} D_{aK} D_{8K} \rangle_{p \uparrow}
 \cdot \tilde{s} (x) ,
\end{eqnarray}
where $\tilde{e} (x)$ is defined in (\ref{ga1ew}), while
$\tilde{f}(x)$ and $\tilde{s}(x)$ are given by
\begin{eqnarray}
 \tilde{f} (x) \!\! &=& \!\! - M_N \frac{N_c}{4 I_1}
 \left( \sum_{m = all, n \leq 0} - \sum_{m = all, n > 0} \right)
 \frac{1}{E_m - E_n}
 \langle n \vert \gamma^0 \vert m \rangle
 \langle m \vert \lambda_3 (\gamma_5 + \Sigma_3) 
 \delta_n \vert n \rangle ,
 \nonumber \\
 \tilde{s} (x) \!\! &=& \!\!- M_N \frac{N_c}{4 I_2}
 \left( \sum_{m = all, n \leq 0} - \sum_{m = all, n > 0} \right)
 \frac{1}{E_m - E_n}
 \langle n \vert \lambda_4 \gamma^0 \vert m \rangle
 \langle m \vert \lambda_4 (\gamma_5 + \Sigma_3) 
 \delta_n \vert n \rangle ,
 \ \ \ \ \ 
\end{eqnarray}

Next we turn to the kinematical $\Delta m_s$ correction,
which originates from the first order correction with respect
to $\Delta m_s$ in the collective quantization rule (\ref{qrulems}).
Putting this rule into the operator
$\Omega$ contained in (\ref{omega1}), we are led to a simple rule for
obtaining the kinematical $\Delta m_s$ correction to
$O_{kin}^{(\Delta m_s)}$, i.e.
\begin{eqnarray}
 \frac{J_i}{2 I_1} \ &\longrightarrow& \ \frac{1}{2} \,
 \frac{2}{\sqrt{3}} \,
 \Delta m_s \,\frac{K_1}{I_1} \,D_{8i} , \\
 \frac{J_K}{2 I_2} \ &\longrightarrow& \ \frac{1}{2} \,
 \frac{2}{\sqrt{3}} \,
 \Delta m_s \,\frac{K_2}{I_2} \,D_{8K} .
\end{eqnarray}
Taking care of the fact that the collective operator contained in
$\tilde{O}$ commute with $D_{8i}$ as well as $D_{8K}$, we therefore
obtain
\begin{eqnarray}
 O_{kin}^{(\Delta m_s)} &=& - D_{8i} \cdot 
 \frac{2}{\sqrt{3}} \Delta m_s
 \frac{K_1}{I_1} M_N \frac{N_c}{4}
 \left( \sum_{m = all, n \leq 0} - \sum_{m = all, n > 0} \right)
 \nonumber \\
 &\,& \times \,\frac{1}{E_m - E_n} \,\left[
 \langle n \vert \bar{O} \delta_n \vert m \rangle
 \langle m \vert \lambda_i \vert n \rangle + 
 \langle n \vert \lambda_i \vert m \rangle
 \langle m \vert \bar{O} \delta_n \vert n \rangle 
 \right], \nonumber \\
 &\,& + \,D_{8i} \cdot \frac{2}{\sqrt{3}} \Delta m_s \frac{K_1}{I_1}
 \frac{N_c}{4} \frac{d}{dx}
 \left( \sum_{n \leq 0} - \sum_{n > 0} \right)
 \langle n \vert \lambda_i \bar{O} \delta_n \vert n \rangle ,
 \label{kinmssng}
\end{eqnarray}
for the flavor singlet distributions in which $\tilde{O}_a = A^\dagger
\lambda_0 A \bar{O} = \bar{O}$. On the other hand, the flavor
nonsinglet part becomes
\begin{eqnarray}
 O_{kin}^{(\Delta m_s)} &=& - D_{ab} D_{8i} \frac{2}{\sqrt{3}}
 \Delta m_s \frac{K_1}{I_1} M_N \frac{N_c}{4}
 \left( \sum_{m = all, n \leq 0} - \sum_{m = all, n > 0} \right)
 \frac{1}{E_m - E_n} \nonumber \\
 &\,& \hspace{10mm} \times \,\left[
 \langle n \vert \lambda_b \bar{O} \delta_n \vert m \rangle
 \langle m \vert \lambda_i \vert n \rangle + 
 \langle n \vert \lambda_i \vert m \rangle
 \langle m \vert \lambda_b \bar{O} \delta_n \vert n \rangle
 \right] \nonumber \\
 &\,& - D_{ab} D_{8K} \frac{2}{\sqrt{3}}
 \Delta m_s \frac{K_2}{I_2} M_N \frac{N_c}{4}
 \left( \sum_{m = all, n \leq 0} - \sum_{m = all, n > 0} \right)
 \frac{1}{E_m - E_n} \nonumber \\
 &\,& \hspace{10mm} \times \,\left[
 \langle n \vert \lambda_b \bar{O} \delta_n \vert m \rangle
 \langle m \vert \lambda_K \vert n \rangle + 
 \langle n \vert \lambda_K \vert m \rangle
 \langle m \vert \lambda_b \bar{O} \delta_n \vert n \rangle
 \right] \nonumber \\
 &\,& + \,D_{ab} D_{8i} \frac{2}{\sqrt{3}} \Delta m_s \frac{K_1}{I_1}
 \frac{N_c}{4} \frac{d}{dx} 
 \left( \sum_{n \leq 0} - \sum_{n > 0} \right)
 \langle n \vert \frac{1}{2} \{ \lambda_b \bar{O}, \lambda_i \}
 \vert n \rangle \nonumber \\
 &\,& + \,D_{ab} D_{8K} \frac{2}{\sqrt{3}} \Delta m_s \frac{K_2}{I_2}
 \frac{N_c}{4} \frac{d}{dx} 
 \left( \sum_{n \leq 0} - \sum_{n > 0} \right)
 \langle n \vert \frac{1}{2} \{ \lambda_b \bar{O}, \lambda_K \}
 \vert n \rangle . \label{kinmsns}
\end{eqnarray}
Here, the last two terms of the above equation are rewritten by using
the relations,
\begin{eqnarray}
 \sum_{M(n)} \langle n \vert \{ \lambda_b, \lambda_i \} 
 \bar{O} \delta_n
 \vert n \rangle
 \ = \ 
 2 \delta_{bi} \sum_{M(n)} \langle n \vert \bar{O} 
 \delta_n \vert n \rangle + 
 \frac{2}{\sqrt{3}} \delta_{b8} \delta_{i3} \sum_{M(n)}
 \langle n \vert \lambda_3 \bar{O} \delta_n \vert n \rangle ,
 \label{identb1}
\end{eqnarray}
and
\begin{eqnarray}
 \sum_{M(n)} \langle n \vert \{ \lambda_b, \lambda_K \} 
 \bar{O} \delta_n \vert n \rangle
 \ = \ 
 \delta_{bK} \sum_{M(n)} \langle n \vert \bar{O} 
 \delta_n \vert n \rangle + 
 2 \delta_{bK} d_{3KK} \sum_{M(n)}
 \langle n \vert \lambda_3 \bar{O} \delta_n \vert n \rangle ,
 \label{identb2}
\end{eqnarray}
which will be proved in Appendix B.
We thus get for the flavor nonsinglet case
\begin{eqnarray}
 O_{kin}^{(\Delta m_s)} &=& - D_{ab} D_{8i} \frac{2}{\sqrt{3}}
 \Delta m_s \frac{K_1}{I_1} M_N \frac{N_c}{4}
 \left( \sum_{m = all, n \leq 0} - \sum_{m = all, n > 0} \right)
 \frac{1}{E_m - E_n} \nonumber \\
 &\,& \hspace{10mm} \times \,\left[
 \langle n \vert \lambda_b \bar{O} \delta_n \vert m \rangle
 \langle m \vert \lambda_i \vert n \rangle + 
 \langle n \vert \lambda_i \vert m \rangle
 \langle m \vert \lambda_b \bar{O} \delta_n \vert n \rangle
 \right] \nonumber \\
 &\,& - \,D_{ab} D_{8K} \frac{2}{\sqrt{3}}
 \Delta m_s \frac{K_2}{I_2} M_N \frac{N_c}{4}
 \left( \sum_{m = all, n \leq 0} - \sum_{m = all, n > 0} \right)
 \frac{1}{E_m - E_n} \nonumber \\
 &\,& \hspace{10mm} \times \,\left[
 \langle n \vert \lambda_b \bar{O} \delta_n \vert m \rangle
 \langle m \vert \lambda_K \vert n \rangle + 
 \langle n \vert \lambda_K \vert m \rangle
 \langle m \vert \lambda_b \bar{O} \delta_n \vert n \rangle
 \right] \nonumber \\
 &\,& + \,\frac{2 \Delta m_s}{\sqrt{3}} \frac{K_1}{I_1}
 D_{ai} D_{8i} \frac{N_c}{4} \frac{d}{dx}
 \left( \sum_{n \leq 0} - \sum_{n > 0} \right)
 \langle n \vert \bar{O} \delta_n \vert n \rangle \nonumber \\
 &\,& + \,\frac{\Delta m_s}{\sqrt{3}} \frac{K_2}{I_2}
 D_{aK} D_{8K} \frac{N_c}{4} \frac{d}{dx}
 \left( \sum_{n \leq 0} - \sum_{n > 0} \right)
 \langle n \vert \bar{O} \delta_n \vert n \rangle \nonumber \\
 &\,& + \,\frac{2 \Delta m_s}{3} \frac{K_1}{I_1}
 D_{a8} D_{83} \frac{N_c}{4} \frac{d}{dx}
 \left( \sum_{n \leq 0} - \sum_{n > 0} \right)
 \langle n \vert \lambda_3 \bar{O} 
 \delta_n \vert n \rangle \nonumber \\
 &\,& + \,\frac{2 \Delta m_s}{\sqrt{3}} \frac{K_2}{I_2}
 \sum_{K=4}^7 d_{3KK} D_{aK} D_{8K} \frac{N_c}{4} \frac{d}{dx}
 \left( \sum_{n \leq 0} - \sum_{n > 0} \right)
 \langle n \vert \lambda_3 \bar{O} \delta_n \vert n \rangle .
\end{eqnarray}

Let us first consider the unpolarized case. From the general
formula (\ref{kinmssng}), it is easy to see that the kinematical
$\Delta m_s$ correction to the flavor singlet unpolarized
distribution identically vanishes, i.e.
\begin{equation}
 q^{(0)} (x ; \Delta m_s^{kin}) = 0 .
\end{equation}
On the other hand, by using the identities
\begin{eqnarray}
 \sum_{M(n)} \langle n \vert (1 + \gamma^0 \gamma^3)
 \delta_n \vert n \rangle
 &=& \frac{1}{3} \sum_{j=1}^3 \sum_{m = all, M(n)}
 \langle n \vert \lambda_j \vert m \rangle
 \langle m \vert \lambda_j (1 + \gamma^0 \gamma^3) 
 \delta_n \vert n \rangle
 \nonumber \\
 &=& 2 \sum_{m = all, M(n)}
 \langle n \vert \lambda_4 \vert m \rangle
 \langle m \vert \lambda_4 (1 + \gamma^0 \gamma^3) 
 \delta_n \vert n \rangle ,
\end{eqnarray}
the kinematical $\Delta m_s$ correction to the flavor nonsinglet
unpolarized distribution can be expressed in the form :
\begin{eqnarray}
 q^{(a)} (x ; \Delta m_s^{kin}) &=& 
 - \frac{2 \Delta m_s I_1}{\sqrt{3}} \,\frac{K_1}{I_1} \,
 \langle \sum_{i=1}^3 \{ D_{ai}, D_{8i} \} \rangle_p 
 \cdot k_1 (x) \nonumber \\
 &\,& 
 - \frac{2 \Delta m_s I_2}{\sqrt{3}} \,\frac{K_2}{I_2} \,
 \langle \sum_{i=4}^7 \{ D_{aK}, D_{8K} \} \rangle_p
 \cdot k_2 (x) .
\end{eqnarray}
Here $k_1 (x)$ and $k_2 (x)$ are the same functions as appeared
in (\ref{k1k2}).

The kinematical $\Delta m_s$ correction to the flavor singlet
longitudinally polarized distribution can similarly be evaluated as
\begin{eqnarray}
 \Delta q^{(0)} (x ; \Delta m_s^{kin}) &=&
 - {\langle D_{83} \rangle}_{p \uparrow} \frac{4 \Delta m_s}{\sqrt{3}}
 \frac{K_1}{I_1} M_N \frac{N_c}{4}
 \left( \sum_{m = all, \leq 0} - \sum_{m = all, n > 0} \right)
 \nonumber \\
 &\,& \hspace{20mm} \times \frac{1}{E_m - E_n}
 \langle n \vert \lambda_3 \vert m \rangle
 \langle m \vert (\gamma_5 + \Sigma_3) \delta_n \vert n \rangle
 \nonumber \\
 &\,&
 + {\langle D_{83} \rangle}_{p \uparrow} \frac{4 \Delta m_s}{\sqrt{3}}
 \frac{K_1}{I_1} \frac{N_c}{8} \frac{d}{dx}
 \left( \sum_{m = all, \leq 0} - \sum_{m = all, n > 0} \right)
 \nonumber \\
 &\,& \hspace{30mm} \times 
 \langle n \vert \lambda_3 \vert m \rangle
 \langle m \vert (\gamma_5 + \Sigma_3) \delta_n \vert n \rangle
 \nonumber \\
 &=& \ \ \frac{4 \Delta m_s I_1}{\sqrt{3}} \,\frac{K_1}{I_1} \,\,
 {\langle D_{83} \rangle}_{p \uparrow} \cdot e(x) ,
\end{eqnarray}
with $e(x)$ defined before in (\ref{ga1e}).
For the flavor nonsinglet piece,
we obtain
\begin{eqnarray}
 q^{(a)} (x ; \Delta_s^{kin})
 \!\! &=& \!\! - \frac{4 \Delta m_s}{3} \,\frac{K_1}{I_1} \,
 {\langle D_{a8} D_{83} \rangle}_{p \uparrow} \,M_N \frac{N_c}{4}
 \left( \sum_{m = all, n \leq 0} - \sum_{m = all, n > 0} \right)
 \nonumber \\
 &\,& \hspace{15mm} \times \,\frac{1}{E_m - E_n} \,
 \langle n \vert \lambda_3 \vert m \rangle
 \langle m \vert (\gamma_5 + \Sigma_3) \delta_n \vert n \rangle
 \nonumber \\
 &\,& \!\! - \frac{2 \Delta m_s}{\sqrt{3}} \,\frac{K_2}{I_2} \,
 \langle 4 \sum_{K=4}^7 d_{3KK} D_{aK} D_{8K} \rangle_{p \uparrow} \,
 M_N \frac{N_c}{4}
 \left( \sum_{m = all, n \leq 0} - \sum_{m = all, n > 0} \right)
 \nonumber \\
 &\,& \hspace{15mm} \times \frac{1}{E_m - E_n}
 \langle n \vert \lambda_4 \vert m \rangle
 \langle m \vert \lambda_4 (\gamma_5 + \Sigma_3) \delta_n 
 \vert n \rangle \nonumber \\
 &+& \! 
\frac{2 \Delta m_s}{3} \,\frac{K_1}{I_1} \,
 {\langle D_{a8} D_{83} \rangle}_{p \uparrow} \,
 \frac{N_c}{4} \frac{d}{dx}
 \left( \sum_{n \leq 0} - \sum_{n > 0} \right)
 \langle n \vert \lambda_3 (\gamma_5 + \Sigma_3) 
 \delta_n \vert n \rangle
 \nonumber \\
 &+& \!\! \frac{2 \Delta m_s}{\sqrt{3}} \,\frac{K_2}{I_2} \,
 \langle \sum_{K=4}^7 d_{3KK} D_{aK} D_{8K} \rangle_{p \uparrow} \,
 \frac{N_c}{4} \frac{d}{dx}
 \left( \sum_{n \leq 0} - \sum_{n > 0} \right)
 \langle n \vert \lambda_3 (\gamma_5 + \Sigma_3) 
 \delta_n \vert n \rangle . \ \ \ \ \ \ \  
\end{eqnarray}
To rewrite the last two terms, we use the identities
\begin{eqnarray}
 \sum_{M(n)} \langle n \vert \lambda_3 (\gamma_5 + \Sigma_3) \delta_n
 \vert n \rangle
 &=& \sum_{m = all, M(n)} \langle n \vert \lambda_3 \vert m \rangle
 \langle m \vert (\gamma_5 + \Sigma_3) \delta_n \vert n \rangle
 \nonumber \\
 &=& 2 \sum_{m = all, M(n)} \langle n \vert \lambda_4 \vert m \rangle
 \langle m \vert \lambda_4 (\gamma_5 + \Sigma_3) 
 \delta_n \vert n \rangle .
\end{eqnarray}
This enables us to express $q^{(a)} (x ; \Delta m_s^{kin})$ in the
form :
\begin{eqnarray}
 q^{(a)} (x ; \Delta m_s^{kin}) &=& 
 \frac{4 \Delta m_s I_1}{3} \,\frac{K_1}{I_1} \,
 \langle D_{a8} D_{83} \rangle_{p \uparrow} \cdot e(x) \nonumber \\
 &+& \frac{2 \Delta m_s I_2}{\sqrt{3}} \,\frac{K_2}{I_2} \,
 \langle 4 \sum_{K=4}^7 d_{3KK} D_{aK} D_{8K} \rangle_{p \uparrow} 
 \cdot s(x) ,
\end{eqnarray}
with $e(x)$ and $s(x)$ being the functions respectively defined
in (\ref{ga1e}) and (\ref{ga1s}).

It is now convenient to express the dynamical and kinematical
$\Delta m_s$ corrections in a combined form. For the unpolarized
distributions, this gives
\begin{eqnarray}
 q^{(0)} (x ; \Delta m_s^{dyn + kin}) &=& 
 - \frac{4 \Delta m_s I_1}{3} \,
 {\langle 1 - D_{88} \rangle}_{p} \cdot \tilde{k}_0 (x) , \\
 q^{(a)} (x ; \Delta m_s^{dyn + kin}) &=& 
 - \frac{4 \Delta m_s I_1}{3} \,
 {\langle \frac{D_{a8}}{\sqrt{3}} \rangle}_{p} \cdot
 \tilde{k}_0 (x) \nonumber \\
 &\,& + \frac{2 \Delta m_s I_1}{\sqrt{3}} \,
 \langle \sum_{i=1}^3 \{ D_{ai}, D_{8i} \} \rangle_{p} \cdot
 \left[ \tilde{k}_1 (x) - \frac{K_1}{I_1} k_1 (x) \right] \nonumber \\
 &\,& + \frac{2 \Delta m_s I_2}{\sqrt{3}} \,
 \langle \sum_{i=4}^7 \{ D_{aK}, D_{8K} \} \rangle_{p} \cdot
 \left[ \tilde{k}_2 (x) - \frac{K_2}{I_2} k_2 (x) \right] ,
\end{eqnarray}
while, for the longitudinally polarized distributions, we have
\begin{eqnarray}
 \Delta q^{(0)} (x ; \Delta m_s^{dyn + kin}) &=& 
 - \frac{4 \Delta m_s I_1}{\sqrt{3}} \,
 {\langle D_{83} \rangle}_{p \uparrow} \cdot
 \left[ \tilde{e} (x) - \frac{K_1}{I_1} e(x) \right] , \\
 \Delta q^{(a)} (x ; \Delta m_s^{dyn + kin}) &=& 
 \ \frac{4 \Delta m_s I_1}{3} \,
 {\langle D_{83} (1 - D_{88}) \rangle}_{p \uparrow} \cdot \tilde{f} (x)
 \nonumber \\
 &\,& - \frac{4 \Delta m_s I_1}{3} \,
 {\langle D_{a8} D_{83} \rangle}_{p \uparrow} \cdot
 \left[ \tilde{e} (x) - \frac{K_1}{I_1} e(x) \right] \nonumber \\
 &\,& - \frac{2 \Delta m_s I_2}{\sqrt{3}} \,
 \langle 4 \sum_{K=4}^7 d_{3KK} D_{aK} D_{8K} \rangle_{p \uparrow} \cdot \,
 \left[ \tilde{s} (x) - \frac{K_2}{I_2} s(x) \right] .
\end{eqnarray}
We summarize below the necessary matrix elements of
collective operators. For the unpolarized case, we need
\begin{eqnarray}
 \langle \frac{D_{38}}{\sqrt{3}} \rangle_p &=& 
 \frac{1}{30}, \hspace{20mm}
 \langle \frac{D_{88}}{\sqrt{3}} \rangle_p \ = \ 
 \frac{\sqrt{3}}{10}, \\
 \langle \sum_{i=1}^3 \{ D_{3i}, D_{8i} \} \rangle_p &=&
 \frac{2 \sqrt{3}}{45}, \hspace{15mm}
 \langle \sum_{i=1}^3 \{ D_{8i}, D_{8i} \} \rangle_p \ =\ 
 \frac{2}{5}, \\
 \langle \sum_{i=4}^7 \{ D_{3K}, D_{8K} \} \rangle_p &=&
 - \frac{2 \sqrt{3}}{45}, \hspace{10mm}
 \langle \sum_{i=4}^7 \{ D_{8K}, D_{8K} \} \rangle_p \ = \ 
 \frac{6}{5} ,
\end{eqnarray}
while, for the longitudinally polarized case,
\begin{eqnarray}
 \langle D_{33} \rangle_{p \uparrow} &=& - \,\frac{7}{30},
 \hspace{35mm}
 \langle D_{83} \rangle_{p \uparrow} \ = \ - \,\frac{\sqrt{3}}{30},\\
 \langle D_{33} \,(1 - D_{88}) \rangle_{p \uparrow} &=&
 - \,\frac{13}{90}, \hspace{15mm}
 \langle D_{83} \,(1 - D_{88}) \rangle_{p \uparrow} \ = \
 - \,\frac{\sqrt{3}}{30}, \\
 \langle D_{38} \,D_{83} \rangle_{p \uparrow} &=&
 - \,\frac{1}{45}, \hspace{30mm}
 \langle D_{88} \,D_{83} \rangle_{p \uparrow} \ = \
 0, \\
 \langle 4 \,\sum_{K=4}^7 \,d_{3KK} D_{3K} D_{8K} \rangle &=&
 - \,\frac{22 \sqrt{3}}{135}, \hspace{10mm}
 \langle 4 \,\sum_{K=4}^7 \,d_{3KK} D_{8K} D_{8K} \rangle \ = \
 - \,\frac{2}{15} .
\end{eqnarray}

Because the 1st-moment sum rules for the unpolarized distributions are 
connected with the quark-number conservation laws and since they are
shown to be satisfied at the leading $O(\Omega^{0} + \Omega^{1})$
contributions to the distribution functions, one must check whether 
the above SU(3) symmetry breaking corrections do not destroy these 
fundamental conservation laws. To verify them, we first notice the 
relations,
\begin{eqnarray}
 \int_{-1}^1 \tilde{k}_0 (x) \,d x &=& 0, \\
 \int_{-1}^1 \tilde{k}_1 (x) \,d x &=& \frac{K_1}{I_1}, \\
 \int_{-1}^1 \tilde{k}_2 (x) \,d x &=& \frac{K_2}{I_2}
\end{eqnarray}
with $I_1,I_2$ and $K_1, K_2$ being the basic moments of inertia of the 
soliton defined in (\ref{momi}) $\sim$ (\ref{momk}).
Combining the above relations with 
the similar sum rules for $k_1 (x)$ and $k_2 (x)$, we then find that
\begin{eqnarray}
 \int_{-1}^1 [ \tilde{k}_1 (x) - \frac{K_1}{I_1} k_1 (x) ] \,d x &=& 
 \frac{K_1}{I_1} - \frac{K_1}{I_1} \ = \ 0, \\
 \int_{-1}^1 [ \tilde{k}_2 (x) - \frac{K_2}{I_2} k_2 (x) ] \,d x &=&
 \frac{K_2}{I_2} - \frac{K_2}{I_2} \ = \ 0.
\end{eqnarray}
It is now evident from these relations that 
\begin{eqnarray}   
 \int_{-1}^1 q^{(0)} (x : \Delta m_s^{d y n + k i n} ) \,d x &=& 0, \\
 \int_{-1}^1 q^{(3)} (x : \Delta m_s^{d y n + k i n} ) \,d x &=& 0, \\
 \int_{-1}^1 q^{(8)} (x : \Delta m_s^{d y n + k i n} ) \,d x &=& 0,
\end{eqnarray}
which ensures that there is no contributions from the dynamical plus 
kinematical $\Delta m_s$ corrections to the quark-number sum rules.

Since the mass difference between the $s$- and $u,d$-quarks breaks 
SU(3) symmetry, a baryon state is no longer a member of the pure 
SU(3) representation but it is generally a mixture of several SU(3)
representations. Up to the first order in $\Delta m_s$, it can be 
shown that the proton state is a linear combination of three SU(3) 
representation as
\begin{equation}
 | p \uparrow \rangle \ = \ | 8,p \uparrow \rangle
 \ + \ c^N_{\overline{10}} \,| \overline{10}, p \uparrow \rangle 
 \ + \ c^N_{27} \,| 27, p \uparrow \rangle .
\end{equation} 
Here, the mixing coefficients are given by
\begin{eqnarray}
 c^N_{\overline{10}} 
 &=& - \frac{\sqrt{5}}{15} \left( \alpha + \frac{1}{2} \gamma \right)
 I_2 ,\\
 c^N_{27}
 &=& - \frac{\sqrt{6}}{25} \left( \alpha - \frac{1}{6} \gamma \right)
 I_2 ,
\end{eqnarray}
where
\begin{eqnarray}
 \alpha 
 &=& \left( - \,\frac{\bar{\sigma}}{N_c} + \frac{K_2}{I_2} \right)
 \Delta m_s ,\\
 \gamma &=& 2 \left( \frac{K_1}{I_1} - \frac{K_2}{I_2} \right)
 \Delta m_s ,
\end{eqnarray}
with $\bar{\sigma}$ being the scalar charge of the nucleon given by
\begin{equation}
 \bar{\sigma} \ = \ N_c \sum_{n \leq 0} \,\langle n | 
 \gamma^0 | n \rangle . \label{scharge}
\end{equation}
The representation mixing correction to any nucleon observables can 
therefore be evaluated based on the formula
\begin{eqnarray}
 \langle p \uparrow | \hat{O} | p \uparrow \rangle
 &=& \ \langle 8, p \uparrow | \hat{O} | 8, p \uparrow \rangle
 \nonumber \\
 &+& 2 c^N_{\bar{10}} \,\langle 10,p \uparrow | \hat{O} | 
 8, p \uparrow \rangle
 \nonumber \\
 &+& 2 c^N_{27} \,\langle 27,p \uparrow | \hat{O} | 8, p \uparrow 
 \rangle \nonumber
 \ + \ O((\Delta m_s)^2 ).
\end{eqnarray}
Here, as for the effective operator $\hat{O}$, we take the basic
$O(\Omega^0 + \Omega^1)$ operators, which can be read from
(\ref{unpsing}) and (\ref{unpnsing}) for the unpolarized
distributions, while from (\ref{lgpsing}) and (\ref{lgpnsing}) for
the longitudinally polarized ones.
From (\ref{unpsing}), it is easy to verify that there is no
representation mixing correction to flavor singlet unpolarized
distribution  
\begin{equation}
 q^{(0)} (x : \Delta m_s^{rep} ) = 0 .
\end{equation}
On the other hand, the representation mixing correction to the flavor 
non-singlet distribution is given by
\begin{eqnarray}
 q^{(a)} (x : \Delta m_s^{rep} ) &=& 2 c^N_{\overline{10}} \,
 \{ \langle \overline{10}, p \uparrow | \frac{D_{a 8}}{\sqrt{3}} | 
 8, p \uparrow \rangle \cdot \,f (x) \nonumber \\
 &+& \langle \overline{10}, p \uparrow | \sum_{i = 1}^3 
 \{ D_{8 i}, R_i \} | 8, p \uparrow \rangle \cdot \,k_1 (x) \nonumber \\
 &+& \langle \overline{10}, p \uparrow | \sum_{K = 4}^7 
 \{ D_{a K}, R_K \} | 8, p \uparrow \rangle \cdot \,k_2 (x) \} \nonumber \\
 &+& 2 c^N_{27} \,
 \{ \langle 27, p \uparrow | 
 \frac{D_{a 8}}{\sqrt{3}} | 8, p \uparrow \rangle \cdot \,f (x) \nonumber \\
 &+& \langle 27, p \uparrow | \sum_{i = 1}^3 
 \{ D_{a i}, R_i \} | 8, p \uparrow \rangle \cdot \,k_1 (x) \nonumber \\
 &+& \langle 27, p \uparrow | \sum_{K = 4}^7 
 \{ D_{a K}, R_K \} | 8, p \uparrow \rangle \cdot \,k_2 (x) \} .
\end{eqnarray}
Given below are the matrix elements of the relevant collective operators :
\begin{eqnarray}
 \langle \overline{10}, p \,|\, \frac{D_{38}}{\sqrt{3}} \,
 |\, 8, p \rangle 
 \! &=& \! - \,\frac{1}{6 \sqrt{5}}, \hspace{18mm}
 \langle \overline{10}, p \,|\, \frac{D_{88}}{\sqrt{3}} \,
 |\, 8, p \rangle
 \ = \ \frac{1}{2 \sqrt{15}}, \\
 \langle \overline{10}, p \,|\, \sum_{i = 1}^3 \{ D_{3 i}, R_i \} \,
 |\, 8, p \rangle \! &=& \!\! \frac{1}{2 \sqrt{5}}, \hspace{6mm}
 \langle \overline{10}, p \,|\, \sum_{i = 1}^3 
 \{ D_{8 i}, R_i \} \,|\, 8, p \rangle \ = \ 
 - \,\frac{3}{2 \sqrt{15}}, \ \ \ \\
 \langle \overline{10}, p \,|\, \sum_{K = 4}^7 \{ D_{3 K}, R_K \} \,
 |\, 8, p \rangle \! &=& \! 0, \hspace{10mm}
 \langle \overline{10}, p \,|\, \sum_{K = 4}^7  \{ D_{8 K}, R_K \} \,
 |\, 8,p \rangle \ = \ 0 ,
\end{eqnarray}
and 
\begin{eqnarray}
 \langle 27,p \,|\, \frac{D_{38}}{\sqrt{3}} \,|\, 8, p \rangle
 \! &=& \! \frac{1}{15 \sqrt{6}}, \hspace{18mm}
 \langle 27,p \,|\, \frac{D_{88}}{\sqrt{3}} \,|\, 8, p \rangle
 \ = \ \frac{1}{5 \sqrt{2}}, \\
 \langle 27,p \,|\, \sum_{i = 1}^3 \{ D_{3 i}, R_i \} \,
 |\, 8, p \rangle
 \! &=& \! \frac{1}{15 \sqrt{6}}, \hspace{10mm}
 \langle 27,p \,|\, \sum_{i = 1}^3 \{ D_{8 i}, R_i \} \,
 |\, 8, p \rangle
 \ = \ \frac{1}{5 \sqrt{2}}, \\
 \langle 27,p \,|\, \sum_{K = 4}^7 \{ D_{3 K}, R_K \} \,
 |\, 8, p \rangle
 \! &=& \!\! - \,\frac{4}{15 \sqrt{6}}, \hspace{3mm}
 \langle 27, p \,|\, \sum_{K = 4}^7 \{ D_{8 K}, R_K \} \,
 |\, 8, p \rangle 
 \ = \ - \,\frac{4}{5 \sqrt{2}} . \ \ \ \ \ \ \ 
\end{eqnarray}
Using these, we finally arrive at
\begin{equation}
 q^{(0)} (x : \Delta m_s^{rep} ) = 0 ,
\end{equation}
and
\begin{eqnarray}
 q^{(3)} (x : \Delta m_s^{rep} ) &=& - \,
 \frac{1}{3 \sqrt{5}} \,\,c^N_{\overline{10}} \,\,
 \left( f (x) - 3 k_1 (x) \right)
 \nonumber \\
 &\,& + \,\frac{2}{15 \sqrt{15}} \,\,c^N_{27} \,\,
 \left( f (x) + k_1 (x) - 4 k_2 (x) \right) , \nonumber \\
 q^{(8)} (x : \Delta m_s^{rep} ) &=& + \,
 \frac{1}{\sqrt{15}} \,\,c^N_{\overline{10}} \,\,
 \left( f (x) - 3 k_1 (x) \right)
 \nonumber \\
 &\,& + \,\frac{2}{5 \sqrt{2}} \,\,c^N_{27} \,\,
 \left( f (x) + k_1 (x) - 4 k_2 (x) \right) .
\end{eqnarray}
Remembering the sum rules for $f (x), k_1 (x)$, and $k_2 (x)$ given in
(\ref{f1sum}), (\ref{k1sum}) and (\ref{k2sum}), we can show that 
\begin{eqnarray}
 \int_{-1}^1 q^{(0)} (x : \Delta m_s^{rep} ) \,d x &=& 0, \\
 \int_{-1}^1 q^{(3)} (x : \Delta m_s^{rep} ) \,d x &=& 0, \\ 
 \int_{-1}^1 q^{(8)} (x : \Delta m_s^{rep} ) \,d x &=& 0, 
\end{eqnarray}
which ensures that the quark-number sum rules are intact by the introduction
of the representation mixing $\Delta m_s$ corrections.

Next, we consider the representation mixing correction to the 
longitudinally polarized distributions. The representation mixing 
correction to the flavor singlet distribution is again zero, i.e.
\begin{equation}
 \Delta q^{(0)} (x : \Delta m_s^{rep} ) = 0 ,
\end{equation}
while, for the flavor-nonsinglet distribution, we have
\begin{eqnarray}
 \Delta q^{(a)} (x : \Delta m_s^{rep} ) 
 &=& 2 \,c^N_{\overline{10}} \,\{ \langle \overline{10}, p \uparrow 
 | D_{a 3} | 8, p \uparrow \rangle \cdot (- g(x) - h(x)) \nonumber \\
 &+& \langle \overline{10}, p \uparrow | \,4 \sum_{i = K}^4
 d_{3KK} \frac{1}{2} \{ D_{a K}, J_K \} 
 | 8, p \uparrow \rangle \cdot s(x) \nonumber \\
 &+& \langle \overline{10}, p \uparrow | \,\frac{1}{2}
 \{ D_{a 8}, J_3 \} | 8, p \uparrow \rangle
 \cdot \frac{2}{\sqrt{3}} \,e(x) \} \nonumber \\
 &+& 2 \,c^N_{27} \,\{ \langle 27, p \uparrow | 
 D_{a 3} | 8, p \uparrow \rangle \cdot (- g(x) - h(x)) \nonumber \\
 &+& \langle 27, p \uparrow | \,4 \sum_{K = 4}^7 d_{3KK} \frac{1}{2}
 \{ D_{a K}, J_K \} | 8, p \uparrow \rangle \cdot s(x) \nonumber \\
 &+& \langle 27, p \uparrow | \,\frac{1}{2}
 \{ D_{a 8}, J_3 \} | 8, p \uparrow \rangle 
 \cdot \frac{2}{\sqrt{3}} \,e(x) \} .
\end{eqnarray}
Here we need the following matrix elements :
\begin{eqnarray}
 \langle \overline{10}, p \uparrow |\, D_{33} \,| 8, p \uparrow 
 \rangle 
 &=& -\frac{\sqrt{5}}{30}, \\
 \langle \overline{10}, p \uparrow |\, 
 4 \sum_{K = 4}^7 d_{3 K K} D_{3 K} J_K \,
 | 8, p \uparrow \rangle &=& -\frac{2 \sqrt{5}}{15}, \\
 \langle \overline{10}, p \uparrow |\, D_{38} J_3 \,| 8, p \uparrow
 \rangle 
 &=& -\frac{\sqrt{15}}{60}, \\
 \langle \overline{10}, p \uparrow |\, D_{83} \,| 8, p \uparrow
 \rangle
 &=& \frac{\sqrt{15}}{30}, \\
 \langle \overline{10}, p \uparrow |\, 
 4 \sum_{K = 4}^7 d_{3 K K} D_{8 K} J_K \,
 | 8, p \uparrow \rangle &=& \frac{2 \sqrt{15}}{15}, \\
 \langle \overline{10}, p \uparrow |\, D_{88} J_3 \,| 8, p \uparrow
 \rangle 
 &=& \frac{\sqrt{5}}{20},
\end{eqnarray}
and 
\begin{eqnarray}
 \langle 27, p \uparrow |\, D_{33} \,| 8, p \uparrow \rangle
 &=& -\frac{\sqrt{6}}{270}, \\
 \langle 27, p \uparrow |\, 4 \sum_{K = 4}^7 d_{3 K K} D_{3 K} J_K \,
 | 8, p \uparrow \rangle &=& -\frac{4 \sqrt{6}}{135}, \\
 \langle 27, p \uparrow |\, D_{38} J_3 \,| 8, p \uparrow \rangle 
 &=& \frac{\sqrt{2}}{60}, \\
 \langle 27, p \uparrow |\, D_{83} \,| 8, p \uparrow \rangle 
 &=& - \,\frac{\sqrt{2}}{30}, \\
 \langle 27, p \uparrow |\, 4 \sum_{K = 4}^7 d_{3 K K} D_{8 K} J_K \,
 | 8, p \uparrow \rangle &=& - \,\frac{4 \sqrt{2}}{15}, \\
 \langle 27, p \uparrow |\, D_{88} J_3 \,| 8, p \uparrow \rangle
 &=& \frac{\sqrt{6}}{20},
\end{eqnarray}
Using these relations, we finally obtain 
\begin{eqnarray}
 \Delta q^{(0)} (x : \Delta m_s^{rep} ) &=& 0 , \\
 \Delta q^{(3)} (x : \Delta m_s^{rep} ) 
 &=& + \,\frac{\sqrt{5}}{15} \,\,c^N_{\overline{10}} \,\,
 (g (x) + h (x) - 4 s (x) - e (x)) \nonumber \\
 &\,& + \frac{\sqrt{6}}{135} \,\,c^N_{27} \,\,
 (g (x) + h (x) - 8 s (x) + 3 e (x) ) , \\
 \Delta q^{(8)} (x : \Delta m_s^{rep} ) 
 &=& - \,\frac{\sqrt{15}}{15} \,\,c^N_{\overline{10}} \,\,
 (g (x) + h (x) - 4 s (x) - e (x) ) \nonumber \\
 &\,& + \,\frac{\sqrt{2}}{15} \,\,c^N_{27} \,\,
 (g (x) + h (x) - 8 s (x) + 3 e (x) ) .
\end{eqnarray}

\section{Concluding remarks}

We have developed a path-integral formulation of the flavor SU(3) CQSM
for evaluating quark and antiquark distribution functions in the nucleon.
It has been done so as to take 
over the advantage of the SU(2) model such that the polarization of Dirac-sea
quarks in the hedgehog mean-field is property taken into account. 
This is essential for making reasonable predictions for the hidden strange 
quark distributions in the nucleon, which has totally non-valence character,
as well as the light-flavor sea quark distribution in the nucleon.
The theory as a whole is based on a double expansion in two small parameters.
The one is the expansion in the collective angular velocity operator
$\Omega$ of the rotating soliton, which can also be regarded as a $1 / N_c$
expansion. Another is the perturbation in the strange- and nonstrange-quark 
mass difference, which is also thought to be small as compared with the
typical energy scale of baryon physics.

As for the SU(3) symmetry breaking corrections, we have taken account of 
three possible corrections, named the dynamical correction, kinematical 
correction and the representation mixing correction, which are all linear 
order in the mass parameter $\Delta m_s$. It was emphasized that the 
simultaneous account of the dynamical and the kinematical corrections are 
essential for maintaining the quark number sum rules. 
Unfortunately, we encounter some subtle problem in the evaluation of
the parton distribution functions at the subleading order of $1 / N_c$
expansion, or more concretely, the $O(\Omega^1)$ contribution to the PDF.
It arises from an ordering ambiguity of two collective space operators in
quantization. In the case of SU(2) CQSM, this ambiguity can be avoided if
one adopts a physically plausible time-order keeping quantization
prescription. However, it appears that this particular quantization procedure
is not compatible with the fundamental dynamical assumption of the SU(3) CQSM,
i.e. the embedding of the SU(3) hedgehog followed by the quantization of
soliton rotation in the full SU(3) collective coordinate space.
On the other hand, one can avoid this incompatibility, if one adopts the 
symmetrized ordering of two collective operators before quantization.
The price to pay for it is, however, that one loses phenomenologically
desirable first order rotational correction to some flavor-nonsinglet 
observables, which we know is essential for resolving the long--standing 
$g_A$ problem in the flavor SU(2) version of the CQSM. Undoubtedly, our
understanding of the theoretical aspects of the model is still incomplete
and some more works should be done for clarifying these questions.

\appendix
\section{Proof of equalities (\ref{identa1}) and (\ref{identa2})}

Here, let us prove two identities (\ref{identa1})
and (\ref{identa2}), which we have used in sect.2,
Using the standard SU(3) algebra
\begin{equation}
\{ \lambda_c, \lambda_i \} = \frac{4}{3} \delta_{c i} 
+ 2 d_{c i e} \lambda_e ,
\end{equation}
we proceed as 
\begin{eqnarray}
 &\,& \sum_{M(n)} \,\langle n | \{ \lambda_c, \lambda_i \}
 (\gamma_5 + \Sigma_3 ) \delta_n | n \rangle \nonumber \\
 &=& \sum_{M(n)} \,\langle n | 
 \left( \frac{4}{3} \delta_{c i} + 2 d_{c i e} \lambda_e \right) 
 (\gamma_5 + \Sigma_3) \delta_n | n \rangle \nonumber \\
 &=& 2 \,d_{c i 3} \sum_{M(n)} \langle n | \lambda_3
 (\gamma_5 + \Sigma_3) \delta_n | n \rangle \nonumber \\
 &=& 2 \,d_{338} \cdot \delta_{c 8} \delta_{i 3}
 \sum_{M(n)} \,\langle n | \lambda_3 (\gamma_5 + \Sigma_3) 
 \delta_n | n \rangle \nonumber \\
 &=& \frac{2}{\sqrt{3}} \,\delta_{c 8} \delta_{i 3}
 \sum_{m = all, M(n)} \,\langle n | \lambda_3 | m \rangle
 \langle m | (\gamma_5 + \Sigma_3) \delta_n | n \rangle
\end{eqnarray}
which proves the first identity. To prove the second identity, we 
first notice that
\begin{eqnarray}
 &\,& \sum_{M(n)} \,\langle n | \{ \lambda_c, \lambda_K \} 
 (\gamma_5 + \Sigma_3) \delta_n | n \rangle \nonumber \\
 &=& \sum_{M(n)} \,\langle n | \left( \frac{4}{3} \delta_{c i}
 + 2 d_{c K e} \lambda_e \right) (\gamma_5 + \Sigma_3) 
 \delta_n | n \rangle \nonumber \\
 &=& 2 \,d_{3 c K} \sum_{M(n)} \,\langle n | \lambda_3
 (\gamma_5 + \Sigma_3) \delta_n | n \rangle ,
\end{eqnarray}
Secondly, we can show that
\begin{eqnarray}
 &\,& \!\!\!\! \sum_{m = all, M(n)} \langle n | \lambda_4 | m \rangle
 \langle m | \lambda_4 (\gamma_5 + \Sigma_3) \delta_n | n \rangle
 \nonumber \\
 &=& \sum_{M(n)} \,\langle n | \lambda_4^2 (\gamma_5 + \Sigma_3 )
 \delta_n | n \rangle \nonumber \\
 &=& \sum_{M(n)} \,\langle n | \left( \frac{2}{3} - 
 \frac{1}{2 \sqrt{3}}
 \lambda_8 + \frac{1}{2} \lambda_3 \right)
 (\gamma_5 + \Sigma_3 ) \delta_n | n \rangle \nonumber \\
 &=& \frac{1}{2} \sum_{M(n)} \,
 \langle n | \lambda_3 (\gamma_5 + \Sigma_3 )
 \delta_n | n \rangle .
\end{eqnarray}
Combining the above two equations, we therefore obtain
\begin{eqnarray}
 &\,& \!\!\! \sum_{M(n)} \,\langle n | \{ \lambda_c, \lambda_K \}
 (\gamma_5 + \Sigma_3 ) \delta_n | n \rangle \nonumber \\
 &=& 4 \,d_{3 c K} \sum_{m = all, M(n)}
 \langle n | \lambda_4 | m \rangle
 \langle m | \lambda_4 (\gamma_5 + \Sigma_3 ) \delta_n | n \rangle ,
\end{eqnarray}
which proves the second identity. 

\vspace{-2mm}
\section{Proof of equalities (\ref{identb1}) and (\ref{identb2})}

Here, we will prove the identities (\ref{identb1}) and (\ref{identb2})
used in sect.2. Utilizing the generalized hedgehog symmetry together
with the standard SU(3) algebra, we can proceed as follows :
\begin{eqnarray}
 &\,& \sum_{M(n)} \langle n \vert \,\{ \lambda_b, \lambda_i \} 
 \bar{O} \delta_n \,\vert n \rangle \nonumber \\
 &=& \sum_{M(n)} \langle n \vert \,\left( \frac{4}{3} \delta_{b i}
 + 2 d_{b i e} \lambda_e \right) \,
 \bar{O} \delta_n \,\vert n \rangle \nonumber \\
 &=& \sum_{M(n)} \langle n \vert \,\left( \frac{4}{3} \delta_{b i}
 + 2 d_{b i 8} \lambda_8 + 2 d_{b i 3} \lambda_3 \right) \,
 \bar{O} \delta_n \,\vert n \rangle \nonumber \\
 &=& \sum_{M(n)} \langle n \vert \,\left( \frac{4}{3} \delta_{b i} + 
 2 \delta_{b i} d_{118} \frac{1}{\sqrt{3}} + 2 d_{833} \lambda_3
 \delta_{b 8} \delta_{i 3} \right) \,\bar{O} \delta_n \,\vert n \rangle
 \nonumber \\
 &=& 2 \,\delta_{b i} \,\sum_{M(n)} 
 \langle n \vert \,\bar{O} \delta_n \,
 \vert n \rangle + \frac{2}{\sqrt{3}} \,\delta_{b 8} \,\delta_{i 3} \,
 \sum_{M(n)} \langle n \vert \,\lambda_3 \bar{O} 
 \delta_n \,\vert n \rangle ,
\end{eqnarray}
where the index $i$ runs from 1 to 3. This proves the first identity
(\ref{identb1}). Similarly, for the second case in which $K$ runs
from 4 to 7, we can show that
\begin{eqnarray}
 &\,& \sum_{M(n)} \langle n \vert \,\{ \lambda_b, \lambda_K \} 
 \bar{O} \delta_n \,\vert n \rangle \nonumber \\
 &=& \sum_{M(n)} \langle n \vert \,\left( \frac{4}{3} \delta_{b K}
 + 2 d_{b K e} \lambda_e \right) \,
 \bar{O} \delta_n \,\vert n \rangle \nonumber \\
 &=& \sum_{M(n)} \langle n \vert \,\left( \frac{4}{3} \delta_{b K}
 + 2 d_{b K 8} \lambda_8 + 2 d_{b K 3} \lambda_3 \right) \,
 \bar{O} \delta_n \,\vert n \rangle \nonumber \\
 &=& \sum_{M(n)} \langle n \vert \,\left( \frac{4}{3} \delta_{b K} + 
 2 \delta_{b K} d_{448} \frac{1}{\sqrt{3}} + 2 d_{3KK} \lambda_3
 \delta_{b K} \right) \,\bar{O} \delta_n \,\vert n \rangle
 \nonumber \\
 &=& 2 \,\delta_{b K} \,\sum_{M(n)} 
 \langle n \vert \,\bar{O} \delta_n \,
 \vert n \rangle + 2 \,\delta_{b K} \,d_{3KK} \,
 \sum_{M(n)} \langle n \vert \,\lambda_3 \bar{O} 
 \delta_n \,\vert n \rangle ,
\end{eqnarray}
which proves the second identity (\ref{identb2}).

\begin{acknowledgments}
This work is supported in part by a Grant-in-Aid for Scientific
Research for Ministry of Education, Culture, Sports, Science
and Technology, Japan (No.~C-12640267)
\end{acknowledgments}

\bibliographystyle{unsrt}
\bibliography{su3cqsm2}

\end{document}